\documentclass[final,authoryear,3p,12pt]{elsarticle}

\usepackage{graphicx}
\usepackage[update,prepend]{epstopdf}
\usepackage{booktabs, caption}
\usepackage{natbib}
\usepackage{fancyhdr}
\usepackage{amsmath}
\usepackage{amsfonts}
\usepackage{multirow}
\usepackage{lscape}
\usepackage{hyperref}
\usepackage{color,soul}
\sethlcolor{green}
\usepackage{ifthen}
\usepackage[utf8]{inputenc}
\usepackage{titletoc}
\usepackage{eso-pic}
\definecolor{123}{rgb}{.9,.9,.9}
\hypersetup{
colorlinks=false,
citecolor=blue,
linkbordercolor={1 1 1}, % {.69 .77 .87} % {1 1 1}
citebordercolor={1 1 1},
urlbordercolor={1 1 1}
}

\usepackage{geometry}

\usepackage{amssymb}

\usepackage{amsthm}
\usepackage{xcolor}

\def\barr{\begin{array}}
\def\ear{\end{array}}
\def\earr{\end{array}}
\def\btab{\begin{tabular}}
\def\etab{\end{tabular}}
\def\bbi{\begin{thebibliography}{99}}
\def\ebi{\end{thebibliography}}
\def\bce{\begin{center}}
\def\ece{\end{center}}
\def\bab{\begin{abstract}}
\def\eab{\end{abstract}}
\def\beqa{\begin{eqnarray}}
\def\eeqa{\end{eqnarray}}
\def\beq{\begin{equation}}
\def\eeq{\end{equation}}
\def\ben{\begin{enumerate}}
\def\een{\end{enumerate}}
\def\bdo{\begin{document}}
\def\edo{\end{document}}
\def\bit{\begin{itemize}}
\def\eit{\end{itemize}}
\def\ble{\begin{letter}}
\def\ele{\end{\letter}}
\def\rta{\rightarrow}
\def\lra{\Leftrightarrow}
\def\pll{_\parallel}
\def\per{_\perp}
\def\al{\alpha}
\def\be{{\beta}}
\def\bed{\dot{\beta}}
\def\beh{\hat{\beta}}
\def\ga{{\gamma}}
\def\de{{\delta}}
\def\ep{\epsilon}
\def\ze{\zeta}
\def\et{\eta}
\def\th{\theta}
\def\io{\iota}
\def\ka{{\kappa}}
\def\la{\lambda}
\def\om{\omicron}
\def\rh{\rho}
\def\si{{\sigma}}
\def\sib{\bar{\sigma}}
\def\ta{\tau}
\def\up{\upsilon}
\def\ph{\phi}
\def\ch{\chi}
\def\ps{\psi}
\def\om{{\omega}}
\def\wo{{\omega}_o}
\def\ab{\bar{a}}
\def\bb{\bar{b}}
\def\Tb{\bar{T}}
\def\Jb{\bar{J}}
\def\Kb{\bar{K}}
\def\Sb{\bar{S}}
\def\Ab{\bar{A}}
\def\At{\tilde{A}}
\def\Et{\tilde{E}}
\def\yt{\tilde{y}}
\def\Fb{\bar{F}}
\def\gb{\bar{g}}
\def\xb{\bar{x}}
\def\pb{\bar{p}}
\def\vSb{{\bf \bar{S}}}
\def\vKb{{\bf \bar{K}}}
\def\f{{\bf f}}
\def\ftp{{\bf f}(\th,\ph)}
\def\a{{\bf a}}
\def\at{{\bf a}(t)}
\def\ao{{\bf a}(\om)}
\def\F{{\bf F}}
\def\E{{\bf E}}
\def\B{{\bf B}}
\def\C{{\bf C}}
\def\x{{\bf x}}
\def\z{{\bf z}}
\def\xd{{\dot{\bf x}}}
\def\vd{{\dot{\bf v}}}
\def\xt{{\bf x}(t)}
\def\tx{(t,\x)}
\def\tpxp{(t',\xp)}
\def\txu{(t_1,\x_1)}
\def\txd{(t_2,\x_2)}
\def\k{{\bf k}}
\def\q{{\bf q}}
\def\kp{{\bf k'}}
\def\xp{{\bf x'}}
\def\A{{\bf A}}
\def\H{{\bf H}}
\def\C{{\bf C}}
\def\D{{\bf D}}
\def\I{{\bf I}}
\def\K{{\bf K}}
\def\S{{\bf S}}
\def\J{{\bf J}}
\def\M{{\bf M}}
\def\P{{\bf P}}
\def\V{{\bf V}}
\def\U{{\bf U}}
\def\X{{\bf X}}
\def\R{{\bf R}}
\def\Q{{\bf Q}}
\def\Om{\Omega}
\def\Ph{\Phi}
\def\mn{\mu\nu}
\def\nm{\nu\mu}
\def\CF{{\cal F}}
\def\CH{{\cal H}}
\def\CO{{\cal O}}
\def\CC{{\cal C}}
\def\CP{{\cal P}}
\def\CW{{\cal W}}
\def\CZ{{\cal Z}}
\def\ul{\underline}
\def\aboverel#1\above#2{\mathrel{\mathop{#2}\limits^{#1}}}
\def\beneathrel#1\under#2{\mathrel{\mathop{#2}\limits_{#1}}}
\def\frac#1#2{{#1 \over #2}}
\def\vev#1{{\left\langle{#1}\right\rangle}}
\def\state#1{|#1\rangle}
\def\vac{\state0}
\def\sol{\state s}
\def\div{\nabla\cdot}
\def\divp{\nabla'\cdot}
\def\grad{\nabla}
\def\curl{\nabla\times}
\def\curlp{\nabla'\times}
\def\lap{\nabla^2}
\def\dal{\Box^2}
\def\pde#1#2{\frac{\partial #1}{\partial #2}}
\def\pdes#1#2#3{\lep\frac{\partial #1}{\partial #2}\rip_{#3}}
\def\nn{{\bf n}}
\def\nnz{{\bf n}_0}
\def\np{{\bf n'}}
\def\epb{\bar{\ep}}
\def\thb{\bar{\th}}
\def\scl{{\cal L}}
\def\epo{\ep(\om)}
\def\epop{\ep(\om')}
\def\g{{\bf g}}
\def\dV{{\bf\bar{V}}}
\def\dT{{\bf\bar{T}}}
\def\iniv{\int d^3x\,}
\def\inia{\int d^2x\,}
\def\inivp{\int d^3x'\,}
\def\iniap{\int d^2x'\,}
\def\inik{\int d^3k\,}
\def\inxt{\int d^3x\,dt\,}
\def\inxtp{\int d^3x'dt'}
\def\dvp{d^3x'\,}
\def\dap{d^2x'\,}
\def\inv{\int_V d^3x\,}
\def\invp{\int_V d^3x'\,}
\def\invy{\int_V d^3y\,}
\def\ina{\int_S d^2x\,}
\def\inac{\oint_S d^2x\,}
\def\inacp{\oint_Sd^2x'\,}
\def\inap{\int_S d^2x'\,}
\def\inay{\int_S d^2y\,}
\def\invl{\int_C d{\bf l}}
\def\invlc{\oint_C d{\bf l}}
\def\inlc{\oint_C dl}
\def\inopc{\oint_C d\om'\,}
\def\invv{\int d^3x\,d^3x'\,}
\def\inko{\int d^3k\,d\om}
\def\intmp{\int_{-\infty}^\infty}
\def\intzp{\int_0^\infty}
\def\ftk{\frac1{\sqrt{2\pi}}\int_{-\infty}^\infty dk\,}
\def\ftz{\frac1{\sqrt{2\pi}}\int_{-\infty}^\infty dz\,}
\def\fto{\frac1{\sqrt{2\pi}}\int_{-\infty}^\infty d\om\,}
\def\ftt{\frac1{\sqrt{2\pi}}\int_{-\infty}^\infty dt\,}
\def\stp{\sqrt{2\pi}}

\journal{}

\makeatletter
\def\ps@pprintTitle{%
 \let\@oddhead\@empty
 \let\@evenhead\@empty
 \def\@oddfoot{\footnotesize\itshape\hfill\today}%
 \let\@evenfoot\@oddfoot}
\makeatother

\begin{document}

\begin{frontmatter}

% {\color{red}Measurement of multivariate business cycle synchronization in the European Union}\\[1em]
\title{Business cycle synchronization within the European Union:\\A wavelet cohesion approach\tnoteref{label1}}

\author[ies,utia]{Lubos Hanus}
\author[ies,utia]{Lukas Vacha\corref{correspondingauthor}}
\cortext[correspondingauthor]{Corresponding author} \ead{vachal@utia.cas.cz}
\address[ies]{Institute of Economic Studies, Charles University, Opletalova 21, 110 00, Prague,  CR}
\address[utia]{Institute of Information Theory and Automation, Academy of Sciences of the Czech Republic, Pod Vodarenskou Vezi 4, 182 00, Prague, Czech Republic}

\tnotetext[label1]{We gratefully acknowledge the financial support from the GAUK, No. 366015 and 588314. This article is part of a research initiative launched by the Leibniz Community. The authors are also grateful for funding of part of this research from the European Union Seventh Framework Programme (FP7/2007-2013) under grant agreement No. 619255. Support from the Czech Science Foundation under the P402/12/G097 (DYME - Dynamic Models in Economics) project is also gratefully acknowledged. We would like to thank Mr. Gilles Doufrenot for helpful comments at the 2nd International Workshop on "Financial Markets and Nonlinear Dynamics" in Paris.}

\begin{abstract}
In this paper, we map the process of business cycle synchronization across the European Union. We study this synchronization by applying wavelet techniques, particularly the cohesion measure with time-varying weights. This novel approach allows us to study the dynamic relationship among selected countries from a different perspective than the usual time-domain models. Analyzing monthly data from 1990 to 2014, we show an increasing co-movement of the Visegrad countries with the European Union after the countries began preparing for the accession to the European Union. With particular focus on the Visegrad countries we show that participation in a currency union possibly increases the co-movement. Furthermore, we find a high degree of synchronization in long-term horizons by analyzing the Visegrad Four and Southern European countries' synchronization with the core countries of the European Union.
\end{abstract}

\end{frontmatter}


\section{Introduction}
\label{intro}
One of the most challenging tasks in economics is to identify, understand, and disentangle the factors and mechanisms that impact the dynamics of macroeconomic variables. Many quantitative econometric techniques have been developed to study the regular fluctuations of macroeconomic indicators and business cycles, e.g., \citet{Baxter:1999aa,Hodrick:1981aa,Harding:2002aa}. This article investigates business cycle synchronization over different time horizons. In order to disentangle the desired information, we apply wavelet methodology working in a time-frequency space. The analysis considers the case of the Visegrad Four, both in terms of the interior relationships among its constituent countries and in terms of the relationships established within the framework of the European Union (EU).

It has been more than two decades since the break-up of the Eastern Bloc;\footnote{The Eastern Bloc was generally formed of the countries of the Warsaw Pact (as Central and Eastern European countries) and the Soviet Union.} following its disintegration, these countries began their independent economic and political journeys. While undertaking their economic transformations during this time, the Czech Republic, Hungary, Poland, and Slovakia began discussing mutual cooperation. Despite their originally different levels of economic maturity and development, their willingness and regional proximity led them to establish the Visegrad Four in 1991. One of the chief aims of this group was to help its members to organize their institutions for faster convergence with and integration into the European Union.\footnote{The Visegrad countries also joined the North Atlantic Treaty Organization in 1999 and applied for membership in the European Union in 1995-1996.} In 2004, these four countries became members of the EU, which obliges them to adopt the Euro currency as part of the integration process. One of the concerns of successful integration into the European Economic and Monetary Union (EMU) is business cycle synchronization, which is motivated by the theory of Optimum Currency Area (OCA) \citep{Mundell:1961aa}. A country joining the OCA gives up its individual monetary policy, which requires a level of integration of macroeconomic variables and policies and affects the costs and benefits enjoyed by the nations \citep{De-Haan:2008aa}. The common currency can be beneficial for both new and former countries in terms of trade transaction costs. Otherwise, at the European level, the European Central Bank controls those policies that apply to all member states, which may be counter-cyclical for countries with low business cycle synchronization \citep{Kolasa:2013aa}. On one hand, these policies may create difficulties for those countries. On the other hand, countries with low levels of synchronization may benefit from being members of the OCA because the business cycle synchronization appears as an endogenous criterion. This endogeneity of OCA means that forming a monetary union will make its members more synchronized \citep{Frankel:1998aa}.\footnote{The literature focusing on the evolution and determinants of business cycle synchronization between Central and Eastern European (CEE) countries and the EU is extensive, e.g., \citet{Darvas:2008aa,Artis:2004aa,Backus:1992aa}.}

The literature regarding EU integration -- and particularly that focusing on the economic integration of the CEE countries -- has grown rapidly. \citet{Fidrmuc:2006aa} conduct a meta-analysis of 35 studies involving the synchronization of the EU and CEE countries and find a high level of synchronization between new member states and the EU. However, only Hungary and Poland among the Visegrad countries reached high synchronization. \citet{Artis:2004aa} and \citet{Darvas:2008aa} obtained the same results studying correlations between the business cycles of the EU and Hungary and Poland. In another model, \citet{Jagric:2002aa} also implies that the economic co-movement of Hungary and Poland is high. Analogously, \citet{Bruzda:2011aa} shows that Poland's economic synchronization with the EU rises when intra-EU synchronization is stable. Recently, \citet{Aguiar-Conraria:2011aa} examine the industrial production index of Euro-12 countries\footnote{These group consists of Austria, Belgium, Finland, France, Germany, Greece, Ireland, Italy, Luxembourg, Netherlands, Portugal, and Spain. We analyze this group and the Visegrad countries.} and analyze business cycle synchronizations within the EU framework, while taking into account the distances among regions. These authors show that countries that are closer to one another show higher synchronization. Moreover, the transition countries show high similarity with the EU after 2005. Nevertheless, Slovakia, a member of the euro area, shows little significant cohesion with the EU. With respect to Hungary and the Czech Republic, their business cycles co-move with the EU-12 after 2005. \citet{Jimenez-Rodriguez:2013aa} also find high correlations of CEE countries (except for the Czech Republic) with the EU business cycle. However, contrary to this result, they find that these countries exhibit a lower level of concordance when a factor model is employed. \citet{Crespo-Cuaresma:2013aa} look at second moments of business cycles in the EU and they report a significant convergence since 90s. Furher, they show there is no decrease in the optimality of the currency area after the EU enlargements.

To assess the degree of similarity or synchronization, researchers have searched for appropriate tools to capture the relevant information. One of the most popular tools is the Pearson correlation coefficient, which simply measures the degree of co-movement in a time domain. However, market-based economies are structured over different time horizons. For this reason, researchers began surveying the behavior of such systems at different frequencies corresponding to different time horizons, and the interest in frequency domain measures has grown. \citet{Christiano:2003aa} proposed a model based on a band pass filter that allowed that desired frequencies of time series to be filtered.\footnote{\citet{Lamo:2013aa} offer a short presentation of a filtering methodology applied to the business cycles in the euro area.} Further, \citet{Croux:2001aa} presented a measure of co-movement, the dynamic correlation, based on a spectral analysis, that equals to basic correlation on a band pass filtered time series. Nevertheless, both the time (static) Pearson correlation and the spectral domain dynamic correlation have several caveats. The first loses information about frequency horizons and the latter omits the co-movement dependence in time. The wavelet analysis overcomes such limitations due to its operation in both time and frequency domains \citep{Torrence:1998aa}. Over the past two decades, wavelet applications have been supported by its another advantage, which is the localization of the wavelet basis function in time and its bounded support; hence, the analysis is free from the assumption of covariance-stationarity, from which many filtering methods suffer \citep{Raihan:2005aa}. The literature presents many studies that successfully used wavelets that do not necessitate stationary time series, e.g., \citet{Aguiar-Conraria:2008aa} analysing the evolution of monetary policy in the US, \citet{Vacha:2012aa} studying energy markets relationships, and \citet{Yogo:2008aa} using wavelet analysis to determine peaks and valleys of business cycles that correspond to the definition of the National Bureau of Economic Research. Recently, \citet{Crowley:2015aa} have used wavelet techniques to disentangle the relationship of the real GDP growth at different frequencies.

To capture the co-movement of two or more time series, we use several wavelet measures that overcome the above mentioned caveats and obtain the desired information about time series relationships. For the bivariate analysis, we employ the wavelet coherence described in \citet{Torrence:1998aa} and \citet{Grinsted:2004aa}. Further, while studying the multiple relationships of several time series, we begin with the bivariate measure proposed by \citet{Rua:2010aa}. As \citet{Croux:2001aa} transforms the dynamic correlation to the multivariate measure of cohesion, \citet{Rua:2012aa} extends the wavelet quantity of \citet{Rua:2010aa} to the multivariate case of weighted cohesion. The multivariate case relies on the bivariate measure multiplied by fixed weights that represent a share of the value of each pair among all time series. However, to reflect the dynamics and development of economies, we believe that weights may change over time and should not be rigid. Taking that into account, we propose a new approach considering that the weights in the cohesion measure have a time-varying structure. In emerging or developing countries higher growth has been observed and they may converge faster and get closer to developed countries; hence, the ratio of each pair significantly changes over time as countries evolve.

Using state of the art wavelet methods, we find different levels of co-movement between Visegrad countries and the EU during the 1990-2014 period. The Visegrad countries show strong co-movement with respect to long-term business cycles. The pairwise synchronization of Visegrad countries with Germany appears to be significant for long-term business cycles from 2000 onward. The measure of multivariate co-movement confirms that the Visegrad countries are well-synchronized for business cycles periods of 2-4 years. Similarly, we observe higher cohesion for the 16 European countries for those periods, which becomes stronger after 2000. All countries together show no considerable relationship for periods up to 1 year, which may reflect some short-term policy heterogeneity.

The contribution of this paper is twofold: we propose the novel measure of cohesion with time-varying weights and we have conducted an empirical analysis regarding the Visegrad Four within the framework of the European Union during the past 2 decades. The remainder of the paper is structured as follows. Section \ref{sec:methodology} describes wavelet methodology and introduces the cohesion measure with time-varying weights. Section \ref{sec:data} provides the data description. In section \ref{sec:results}, we analyze the results. Finally, Section \ref{sec:conclusion} concludes.

\section{Methodology}
\label{sec:methodology}

\subsection{Wavelet analysis}

Our analysis aims to address the behavior of the dynamics of a time series at different time horizons. The Fourier analysis is convenient for observing relations at different frequencies, but it requires time series to be stationary and comes at the cost of losing some information of the time series when differencing, for instance.  Many economic time series might be locally stationary or non-stationary. In other words, using the Fourier transform (FT) makes the analysis time-invariant and not suitable to provide information about the dynamics of a process. For this reason, \citet{Gabor:1946aa} developed the short-time Fourier transform (or windowed FT), which is based on applying the Fourier transform on a shorter part of the process. Nonetheless, this approach has shortcomings that arise with fixed time and frequency resolution, and it is impossible to change the resolution at different frequencies. A series of lower or higher frequencies needs lower or higher time resolution, respectively \citep{Gallegati:2008aa}. As the window width is constant, the resolution is limited, especially for low frequencies.

The wavelet transform has been developed to find a better balance between time and frequency resolutions. The wavelet functions used for the decomposition are narrow or wide when we analyze high or low frequencies, respectively \citep{Daubechies:1992aa}. Thus, a wavelet analysis is suitable to research different types of processes using optimal time-frequency resolution in great detail \citep{Cazelles:2008aa}.\footnote{It is possible to use methods of evolutionary spectra of non-stationary time series developed by \citet{Priestley:1965aa}. However, to run a spectrum over time at different frequencies, larger data are required in order to obtain the same quality time resolution as that obtained using wavelet techniques. And while we study short-term fluctuations (such as 2-4 months), we would not obtain such localized information because of the number of necessary observations to start with.} The wavelet analysis is also able to work properly with locally stationary time series \citep{Nason:2000aa,Crowley:2007aa}.

The wavelet transform decomposes a time series using functions called mother wavelets $\psi(t)$ that are functions of translation (time) and dilation (scale) parameters, $\tau$ and $s$, respectively.
In many applications, it is sufficient that the mother wavelet has zero mean, $\int_{-\infty}^{\infty} \psi(t) \mathrm{d}t = 0$, and that the function has a sufficient decay. These two properties cause the function to behave like a wave. Further, the elementary functions called daughter wavelets resulting from a mother wavelet $\psi(\tau)$ are defined as
\begin{equation}
  \psi_{\tau,s}(t) = \frac{1}{\sqrt{|s|}} \psi \left( \frac{t-\tau}{s} \right), ~~~ s, ~\tau \in \mathbf{R}, s \neq 0.
\end{equation}

The continuous wavelet transform (CWT) of a process, $x(t)$, with respect to the wavelet $\psi$,\footnote{For more details regarding the conditions that a mother wavelet must fulfill, see, e.g., \citet{Mallat:1999aa} or \citet{Daubechies:1992aa}.} is defined as a convolution of the given process and the family $\psi_{\tau,s}$,
\beq
W_x (\tau,s) = \int_{-\infty}^{\infty} x(t) \frac{1}{\sqrt{|s|}}\psi^* \left( \frac{t-\tau}{s} \right) \mathrm{d}t,
\label{eq:cwtTransIntegral}
\eeq
where $*$ denotes the complex conjugate. As several wavelet functions are available for CWT, in our analysis we use the Morlet wavelet as the mother wavelet. Use of this wavelet is common in the literature due to its well-localized properties in time and frequency \citep{Aguiar-Conraria:2011ac,Cazelles:2008aa}. The Morlet wavelet is complex, with a real and an imaginary part, which allows us to perform the phase difference analysis. The simple definition of the Morlet wavelet is
\begin{equation}
  \psi(t) = \pi^{-\frac{1}{4}} e^{-i \omega_0 t} e^{\frac{-t^2}{4}}.
\end{equation}
For a complex function with a Gaussian envelope, the Morlet wavelet's parameter $\omega_0$ (wavenumber) is set equal to 6. With this parameter, we find that a relation between wavelet scales and frequencies is inverse, $f \approx \frac{1}{s}$. This simplifies further interpretation of the results \citep{Grinsted:2004aa,Torrence:1998aa}.

A convenient property of the wavelet transform is that one can reconstruct the original time series back from the wavelet transform,
\beq
x(t) = \frac{1}{C_\psi} \int_{-\infty}^{\infty} \left[ \int_{-\infty}^{\infty} \psi_{\tau,s} (t) W_x (\tau,s)  \mathrm{d}\tau \right] \frac{\mathrm{d}s}{s^2},
\eeq
where $C_{\psi}$ comes from the admissibility condition that allows the reconstruction, $C_{\psi} = \int_0^{\infty} \frac{|\Psi (f)|^2 }{f} \mathrm{d}f < \infty$, where $\Psi(\cdot)$ is the Fourier transform of $\psi(\cdot)$.

Defining the single wavelet power spectrum, $|W_x(\tau,s)|^2$, given the wavelet transform, we obtain the measure of the energy of the time series. For a given two time series, $x_i(t)$ and $x_j(t)$, the cross-wavelet transform is defined as the product of their transforms $W_{x_i x_j}(\tau,s) = W_{x_i}(\tau,s) W_{x_j}(\tau,s)^*$, where $*$ denotes the complex conjugate.

\subsection{Wavelet coherence and phase difference}
In our application, we use the wavelet coherence to quantify the pairwise relationship of two time series. This coherence in the Fourier analysis is defined as a measure of the correlation between the spectra of two time series \citep{Cazelles:2008aa}. The coherence derives from the definition of the coherence as its power of two. With two time series, $x_i$ and $x_j$, the wavelet coherence that measures their relationship is defined as \citep{Liu:1994aa}:
\beq
\Gamma(\tau,s) = \frac{W_{x_i x_j}(\tau,s)}{\sqrt{W_{x_i}(\tau,s)W_{x_j}(\tau,s)}}.
\label{eq:coherency}
\eeq

Given that the coherence is as complex a measure as the wavelet powers are, we preferably use the squared wavelet coherence to measure co-movement between two time series, which is given by:
\beq
R^2(\tau,s) = \frac{|S(s^{-1}W_{x_i x_j}(\tau,s))|^2}{S(s^{-1}|W_{x_i}(\tau,s)|^2) \cdot S(s^{-1}|W_{x_j}(\tau,s)|^2)},~~~R^2 \in [0,1],
\eeq
where $S$ is a smoothing function as $S(W) = S_{scale}(S_{time}(W_n(s))$ \citep{Grinsted:2004aa}.\footnote{We use the package developed by \citet{Grinsted:2004aa} to compute the coherence. For further details, please consult \citet{Grinsted:2004aa}.} We smooth the coherence through convolution in both time and frequency domains.

Since the wavelet coherence does not have a theoretical distribution, the testing procedure uses Monte Carlo methods to obtain its significance. We follow \citet{Torrence:1998aa} to assess the statistical significance, which is depicted in figures as a black contour and the level of significance is 5\%.

As we work with a finite-length time series and a Fourier transform assumes cyclical data, we would obtain a wavelet power spectrum containing errors at the beginning and end of the analyzed periods. One solution to these edge effects is to pad both ends of the time series with a sufficient number of zeros. The area affected by zero-padding is called the cone of influence (COI). We indicate the COI in figures as a shaded area having $e$-folding shape.\footnote{We use the Morlet wavelet, and the COI is thus $e^{-2}$-folding.} For more details, consult \citet{Cazelles:2008aa,Torrence:1998aa}.

Moreover, from the cross wavelet transform of two time series, the phase difference provides information regarding the relative position of the two series. The phase difference defined in $[-\pi, \pi]$ has the form:
\beq
\phi_{x_i,x_j} = \tan^{-1} \left( \frac{\Im \{ W_{x_i x_j}(\tau,s)\}}{\Re\{ W_{x_i x_j}(\tau,s)\}}  \right),
\eeq
where $\Im \{W_{x_i x_j}(\tau,s)\}$ and $\Re \{W_{x_i x_j}(\tau,s)\}$ are the imaginary and real parts of a cross wavelet transform, respectively. The two time series are positively correlated if $\phi_{x_i,x_j} \in [-\pi/2, \pi/2]$, otherwise the correlation is negative. Moreover, the first variable, $x_i$, leads the second, $x_j$, if the phase is in $[0, \pi/2]$ and $[-\pi, -\pi/2]$; when in $[-\pi/2, 0]$ and $[\pi/2, \pi]$, the second variable is leading.

The assessment of statistical significance is always critical. According to \citet{Cazelles:2008aa}, bootstrap methods are used to provide the significance of the power spectrum and the cross-spectrum. Those methods are also used for wavelet coherence. However, testing the significance of the phase difference is difficult because there is no ``preferred'' value because the phase may be distributed on the interval of $[-\pi, \pi]$. \citet{Aguiar-Conraria:2014aa} indicate that there are no good statistical tests for the phase difference. They conclude with the support of \citet{Ge:2008aa} that the significance of a phase should be connected with the significance of the power spectrum or coherence. To obtain the confidence intervals, we use classical bootstrap techniques.\footnote{We add 5\% noise to each analyzed series. We do the wavelet analysis 1,000 times in a  Monte Carlo study, then we sort the results and determine the 95\% confidence interval of the phase difference.} We show the coherence and phase difference of two artificial time series:
\begin{equation}
x_t = \sin(t) + \varepsilon,   \quad  t \in [1,1000] %randn(size(t))/50; + u_t
\end{equation}
\begin{equation}
y_t = \left\{
\begin{array}{l l}
    \frac{1}{100}\sin(t) + \varepsilon, & \quad  t \in [1,100] \\
    \sin(t) + \varepsilon, & \quad  t \in [101,350] \\
    \frac{1}{1000}\sin(t-0.01) + \varepsilon, & \quad  t \in [351,605] \\
    \sin(t+\frac{\pi}{2}) + \varepsilon, & \quad  t \in [606,900] \\
    \frac{1}{100}\sin(t) + \varepsilon, & \quad  t \in [901,1000]
  \end{array} \right.  . %+ v_t,
\end{equation}
\begin{figure}[ht]
\centering
  \captionsetup{width=\textwidth}
  \includegraphics[width=\textwidth, height=10cm]{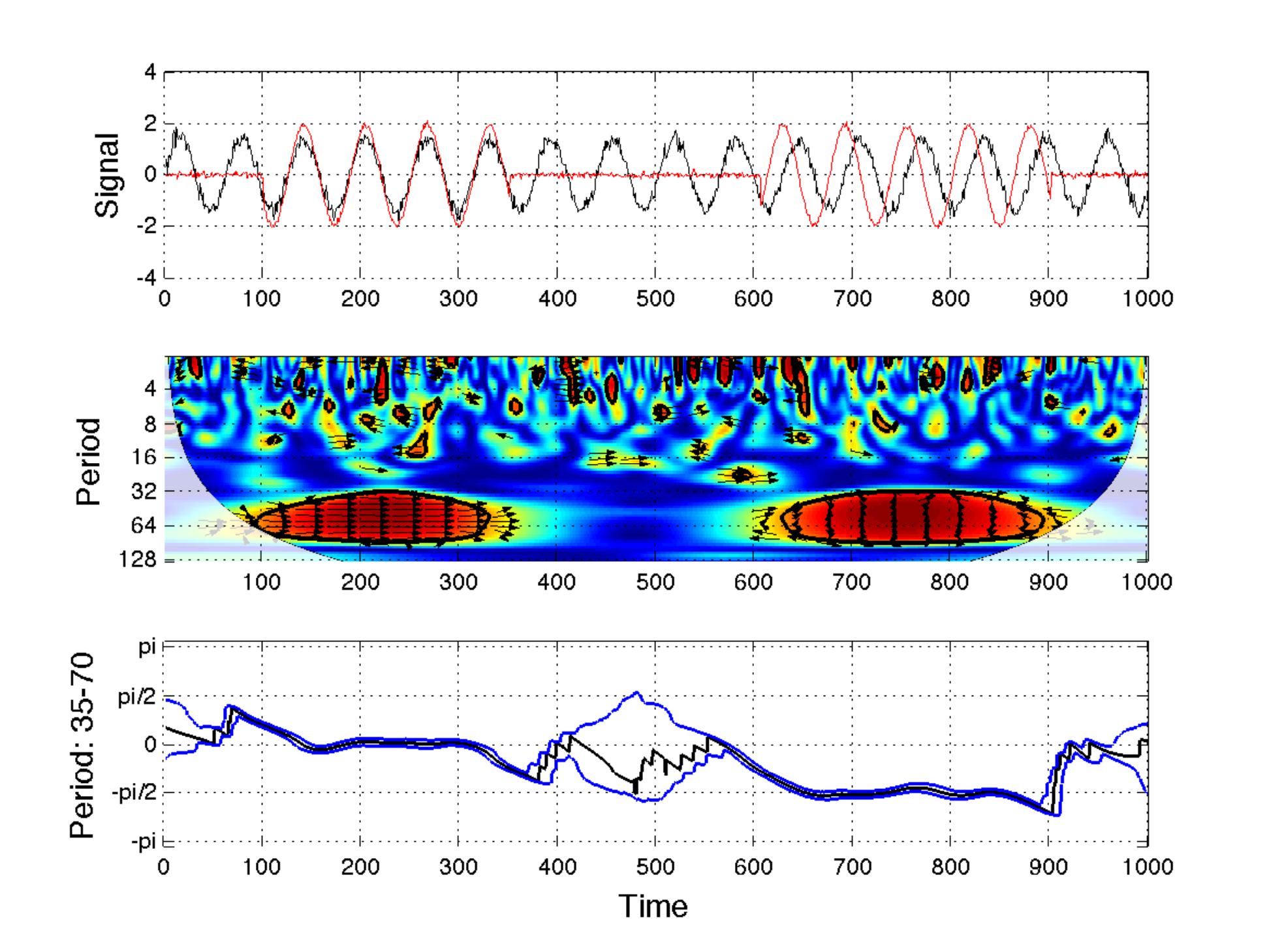}
  \caption[The phase difference of simulated time series]{Artificial time series (top), their wavelet coherence (middle), and the phase difference of the time series (bottom), with 95\% confidence interval.} % The shaded area is the cone of influence.
  \label{fig:simPhase}
\end{figure}

We observe that when the coherence is high and significant, the confidence interval of the phase difference is narrow, see Fig.~\ref{fig:simPhase}. By contrast, for observations around 400-600, we have the sine function with high amplitude and a very noisy time series, which does not resemble the sine in this scale. The phase difference of these two is unstable and the confidence interval at some points covers the interval $[-\pi/2, \pi/2]$. This does not provide relevant information about the phase of the two time series.

%%%%%
% Since the dynamic correlation provides the information over frequencies and disregards the time information, \citet{Rua:2010aa} proposes a wavelet-based measure of co-movement of two time series. This co-movement measure is similar to the wavelet coherency (Eq.\ref{eq:coherency}) but defined as a real number on $[-1,1]$, which inclines more to the dynamic correlation allowing for negative relationship between variables. The measure of co-movement in the time and frequency, $\rho_{x_i x_j}(\tau,s)$, is given by
In order to capture also the negative correlation, Rua (2010) proposes co-movement measure defined as a real number on [-1,1]. It is based on wavelet coherency (Eq. 5), but in the nominator only the real part of a wavelet cross-spectra is used. The measure is given by
\beq
\rho_{x_i x_j}(\tau,s) = \frac{\Re (W_{x_i x_j}(\tau,s))}{\sqrt{|W_{x_i}(\tau,s)|^2 |W_{x_j}(\tau,s)|^2}},
\label{eq:ruaRealCor}
\eeq
where $\Re(W_{x_i x_j}(\tau,s))$ is the real part of the cross-wavelet spectrum of two time series and has the squared root of two power spectra of the given time series in the denominator. Using this measure, \citet{Rua:2012aa} developed a wavelet-based measure of cohesion in the time-frequency.

\begin{figure}[ht]
  \begin{minipage}[t]{0.45\textwidth}
    \includegraphics[width=\textwidth]{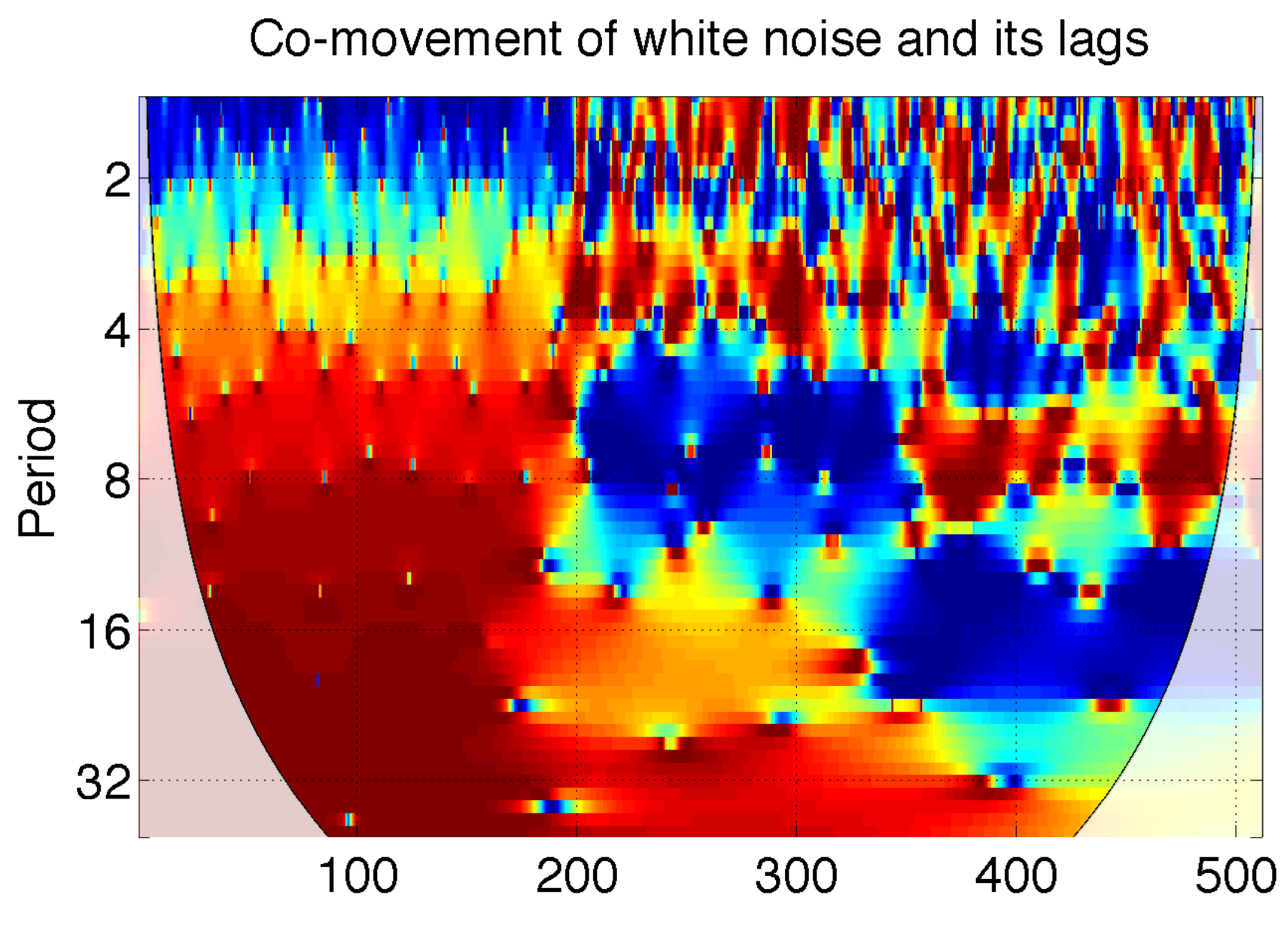}
    \caption{Real wavelet-based measure of co-movement for two series: $a_t = u_t$  for $t=[0,511]$; and $b_t = u_{t-1}$ for $t=[0,200]$, $b_t = u_{t-4}$ for $t=[201,350]$, and $b_t = u_{t-8}$ for $t=[351,511]$.}
    \label{fig:aNoises}
  \end{minipage}
  \hfill
  \begin{minipage}[t]{0.43\textwidth}
    \includegraphics[width=\textwidth]{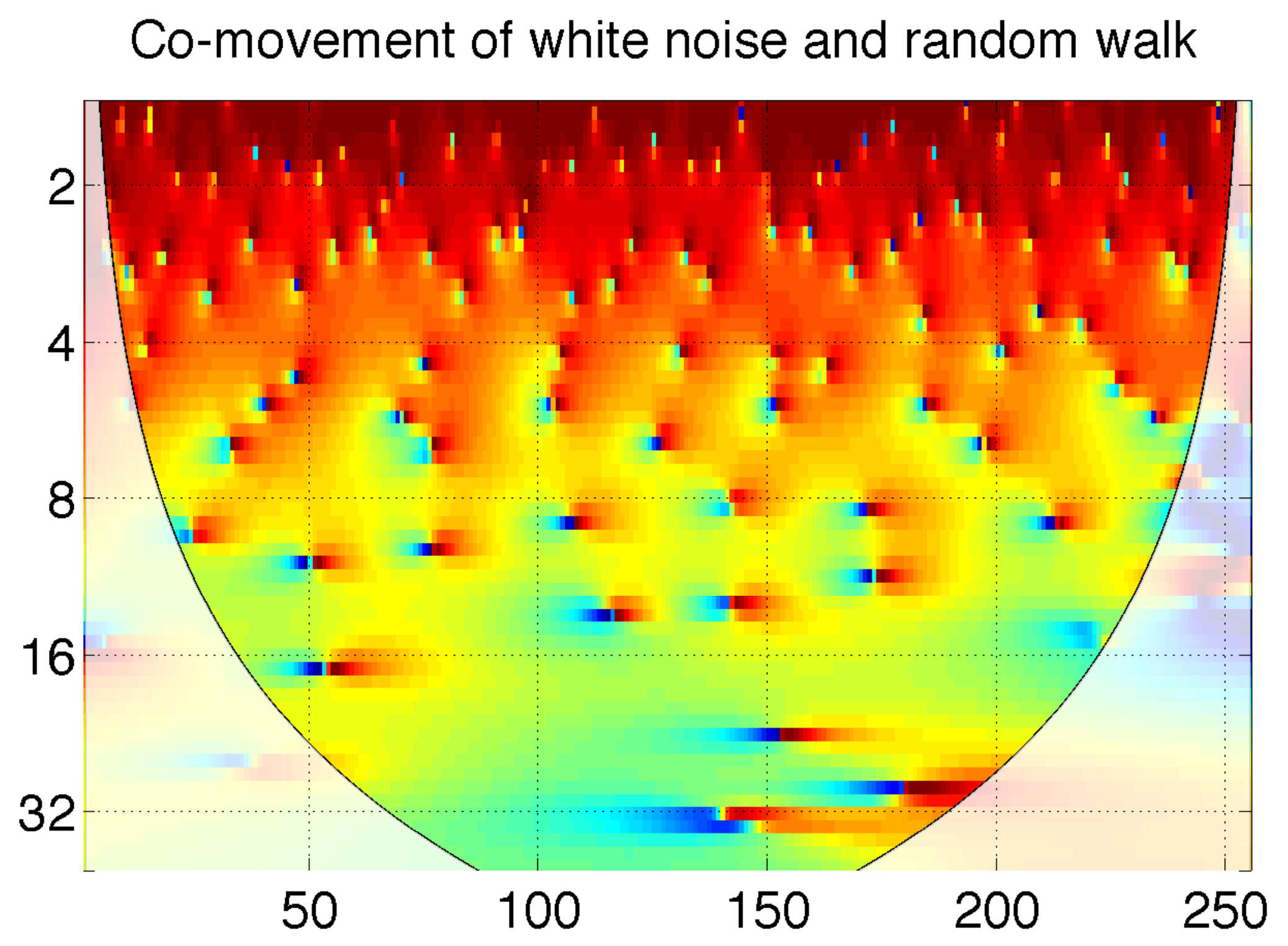}
    \caption{Real wavelet-based measure of co-movement for two series: $a_t = u_t$ and $b_t=b_{t-1}+u_t$.}
    \label{fig:bCumsum}
  \end{minipage}
  \begin{minipage}[t]{0.044\textwidth}
  % \vspace{0.1em}
    \includegraphics[width=\textwidth]{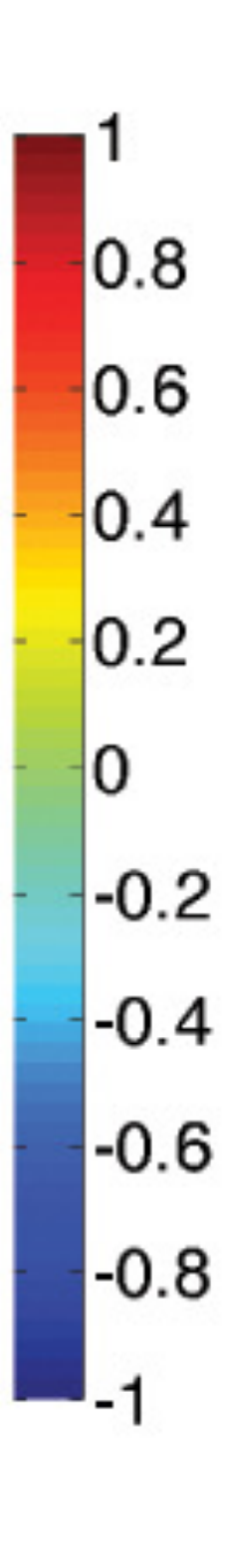}
    %\caption{}
    %\label{fig:1}
  \end{minipage}
\label{fig:simulated}
\end{figure}
%%%%%
We demonstrate usefulness of the real wavelet-based measure (Eq.~\ref{eq:ruaRealCor}) that captures the dynamics in time-frequency of two artificial time series in two particular cases. We show the cohesion with constant weights of white noise, $u_t \sim N(0,1)$ and its lagged values, $u_{t-1}$, $u_{t-4}$, and $u_{t-8}$. For the first 200 observations in Fig.~\ref{fig:aNoises}, we see the negative correlation equal to minus one in the shortest period, which changes to a positive correlation equal to one in the long-term. If we averaged this part of Fig.~\ref{fig:aNoises} over time, the obtained result would be same as the dynamic correlation of \citet{Croux:2001aa}.\footnote{The correlation follows a curve from $1$ at zero frequency to $-1$ at frequency equals $\pi$.} In the second part of Fig.~\ref{fig:aNoises}, the series have more lags, $u_{t-4}$ and $u_{t-8}$, whose relationship is negative with the original $u_t$ at longer horizons, which demonstrates the possibility of the well-localized information in the time-frequency plane. In Fig.~\ref{fig:bCumsum}, we plot the dynamic correlation of the white noise $u_t$ with its cumulative sum. Contrary to Fig.~\ref{fig:aNoises} for lagged noises, these two series are positively correlated in the short-term and not-correlated in the long-term.

% \subsection{Multivariate co-movement measurement}
\subsection{Wavelet cohesion with time-varying weights}
Many co-movement measures from the time or frequency domain rely on the bivariate correlation. \citet{Croux:2001aa} propose a powerful tool for studying the relationship of multiple time series. This measure uses the dynamic pairwise correlation and composes a new measure of cohesion over the frequencies. Let $x_t = (x_{1t} \cdots x_{nt})$ be a multiple time series for $n \geq 2$, then the cohesion measure in the frequency domain is
\begin{equation}
  coh(\lambda) = \frac{\sum_{i \neq j} w_{i} w_j \rho_{x_i x_j}(\lambda)}{\sum_{i \neq j} w_{i} w_j},
\end{equation}
where $\lambda$ is the frequency, $-\pi \leq \lambda \leq \pi$, $w_i$ is the weight associated with time series $x_{it}$, and $coh(\lambda) \in [-1,1]$.

% \subsection{Wavelet cohesion with time-varying weights}
%Second,
In the same manner as \citet{Croux:2001aa} define dynamic correlation in the frequency domain, \citet{Rua:2012aa} take the advantage of the wavelet-based quantity (Eq.~\ref{eq:ruaRealCor}) and define the multivariate weighted measure of cohesion in the time-frequency space. The cohesion is a weighted average, where the weights, $\bar{\omega}_{ij}$, are attached to the pair of series, $(i,j)$, e.g., for two series we have two weights inversely related.
The cohesion exists on the interval of $[-1,1]$ and we have
\beq
coh(\tau, s) = \frac{\sum_{i \neq j} \bar{\omega}_{ij} \rho_{x_i x_j}(\tau,s)}{\sum_{i \neq j} \bar{\omega}_{ij}}.
\label{eq:ruaCoh}
\eeq
Measuring co-movement of multiple time series, the cohesion uncovers important information about common dynamics. However, the fixed weights, e.g., population at some time $\tau$, do not consider that data used for weights may also vary over time. Because the developing or emerging countries may have different speed of development, then the importance of allowing the time-variation of weights appears relevant.

We propose a new approach to map a dynamic multivariate relationship using the time-varying weights in the cohesion measure, which is based on \citet{Rua:2012aa}:
\beq
coh^{TV}(\tau, s) = \frac{\sum_{i \neq j} \omega_{ij}(\tau) \rho_{x_i x_j}(\tau,s)}{\sum_{i \neq j} \omega_{ij}(\tau)}, %\bar{\omega}_{ij}(t)
\label{eq:myCoh}
\eeq
where $\omega_{ij}(\tau)$ is the weight attached to the pair of time series $(i,j)$ at given time $\tau$. Cohesion allows for the use of different types of weights. For example, using GDP as a weight representing the size of an economy, a country with smaller or larger GDP can have smaller or larger effect on a co-movement than other countries, and in the cohesion, this may lead to a greater dissimilarity of co-movement within the group.

\section{Data}
\label{sec:data}
To study business cycle synchronization, we use data of the Index of Industrial Production (IIP) from the database of the Main macroeconomic indicators \citep{OECD:2015aa}.\footnote{Obtained via Federal Reserve Economic Data, \url{https://research.stlouisfed.org/fred2/}} \citet{Fidrmuc:2006aa} cite many studies where the IIP is broadly used in studying business cycle synchronization. The dataset period spans from January 1990 to December 2014 with seasonally adjusted time series. The dataset includes monthly data of 16 EU countries, of which 13 are EMU members (Austria, Belgium, Germany, Greece, Finland, France, Ireland, Italy, Luxembourg, Netherlands, Portugal, Slovakia, Spain) and 3 countries are not (Czech Republic, Hungary, Poland). In the multivariate analysis, we divide the countries into three groups: EU core, Visegrad Four (V4), and PIIGS. Five countries form the PIIGS group: four from Southern Europe (Portugal, Italy, Greece, and Spain) and Ireland. As the EU core, we take Belgium, Germany, Finland, France, Luxembourg, and Netherlands.

For all countries, we also use the data of the Gross Domestic Product (GDP) in current prices  \citep{Statistical-Office-of-the-European-Communities:2016aa}\footnote{Obtained via EUROSTAT, \url{http://ec.europa.eu/eurostat/web/products-datasets/-/namq_10_gdp}, Jan~18,~2016.} and GDP at power purchasing parity (PPP) per capita \citep{World-Bank:2016aa}\footnote{Obtained via The World Bank \url{http://data.worldbank.org/indicator/NY.GDP.PCAP.PP.CD}, Jan~13,~2016.} to weight the industrial production in the multivariate analysis. The wavelet analysis is data demanding, to cover a longer period, we take an advantage of the IIP data that is on monthly basis to represent the economic activity. And while weighting the IIP in the multivariate analysis, we have the GDP data on quarterly basis (GDP in current prices) and on yearly basis (GDP at PPP per capita). This lower frequency samplings are sufficient to have a pronounced impact through time-varying weights at (longer) business cycle frequencies. Our interest is in co-movements longer than a quarter, thus, the quarterly or yearly based weights are acceptable. %The maximum span for analyzed countries of the GDP data is from 1997:Q1 to 2014:Q1 and 1997--2014, respectively.

\section{Results}
\label{sec:results}
\subsection{Bivariate synchronization of the Visegrad countries and the EU} % -- between one another --

\begin{figure}[!ht]
  \begin{minipage}{0.31\textwidth}
    \includegraphics[width=\textwidth]{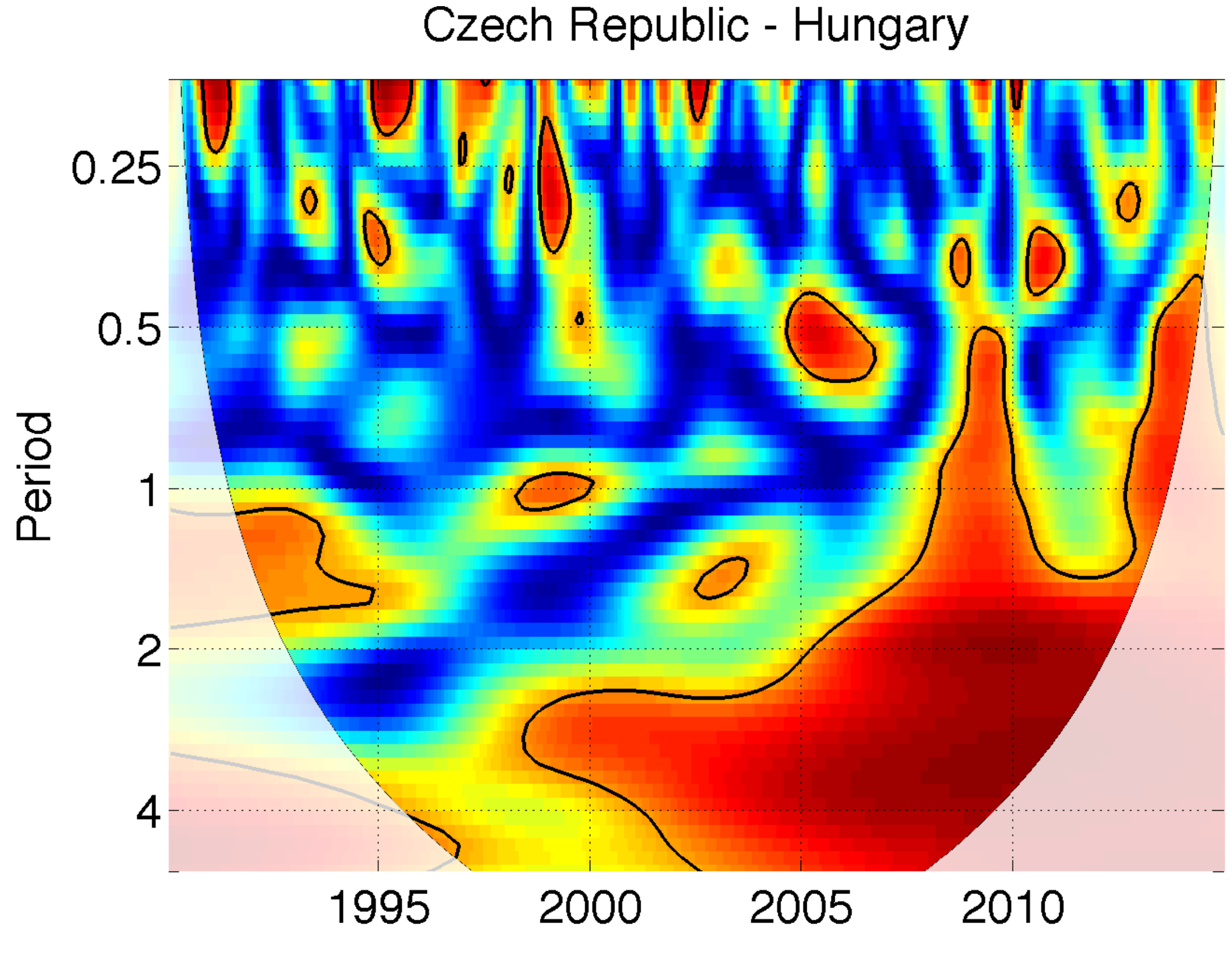}
  \end{minipage}
  \begin{minipage}{0.31\textwidth}
    \includegraphics[width=\textwidth]{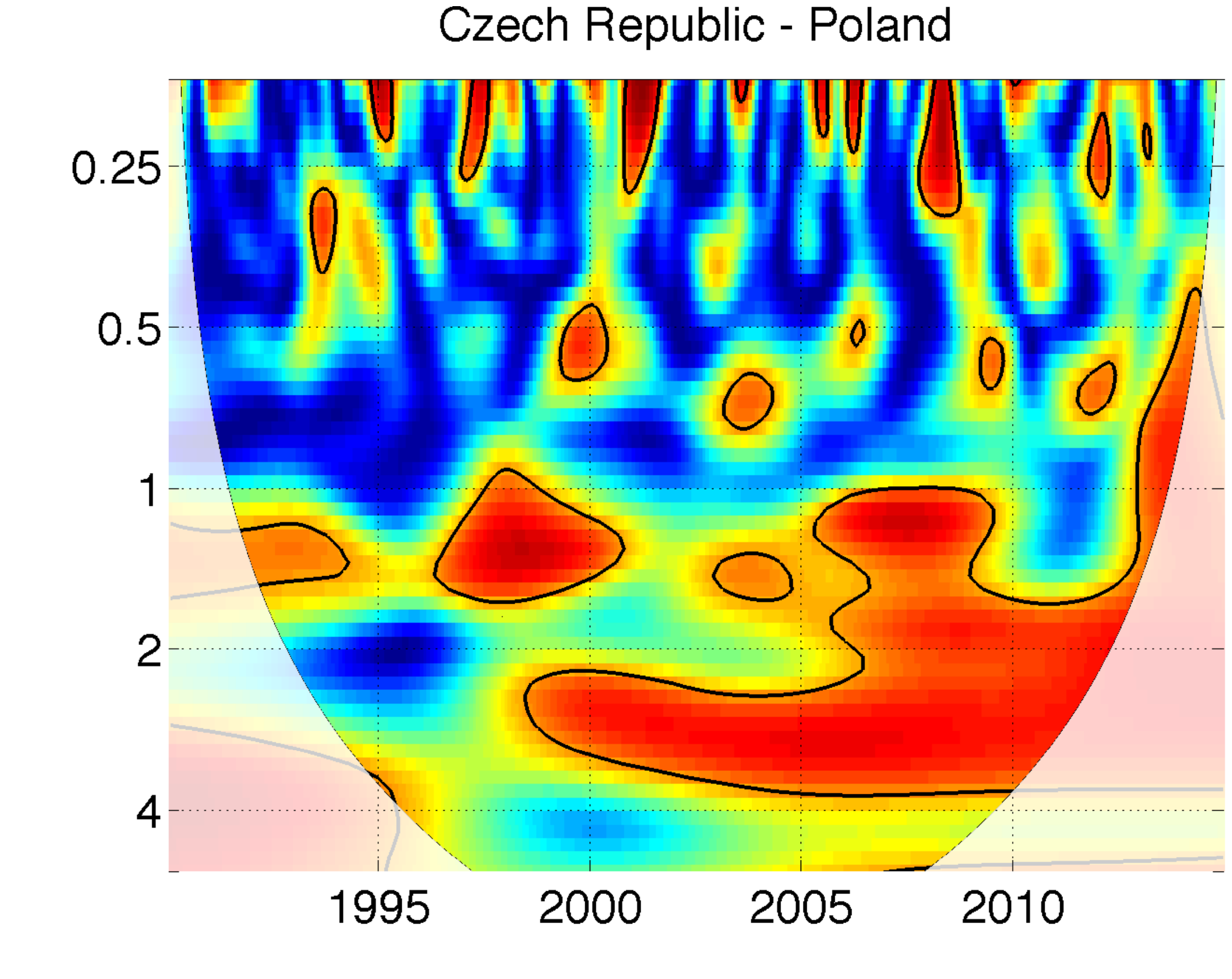}
  \end{minipage}
  \begin{minipage}{0.31\textwidth}
    \includegraphics[width=\textwidth]{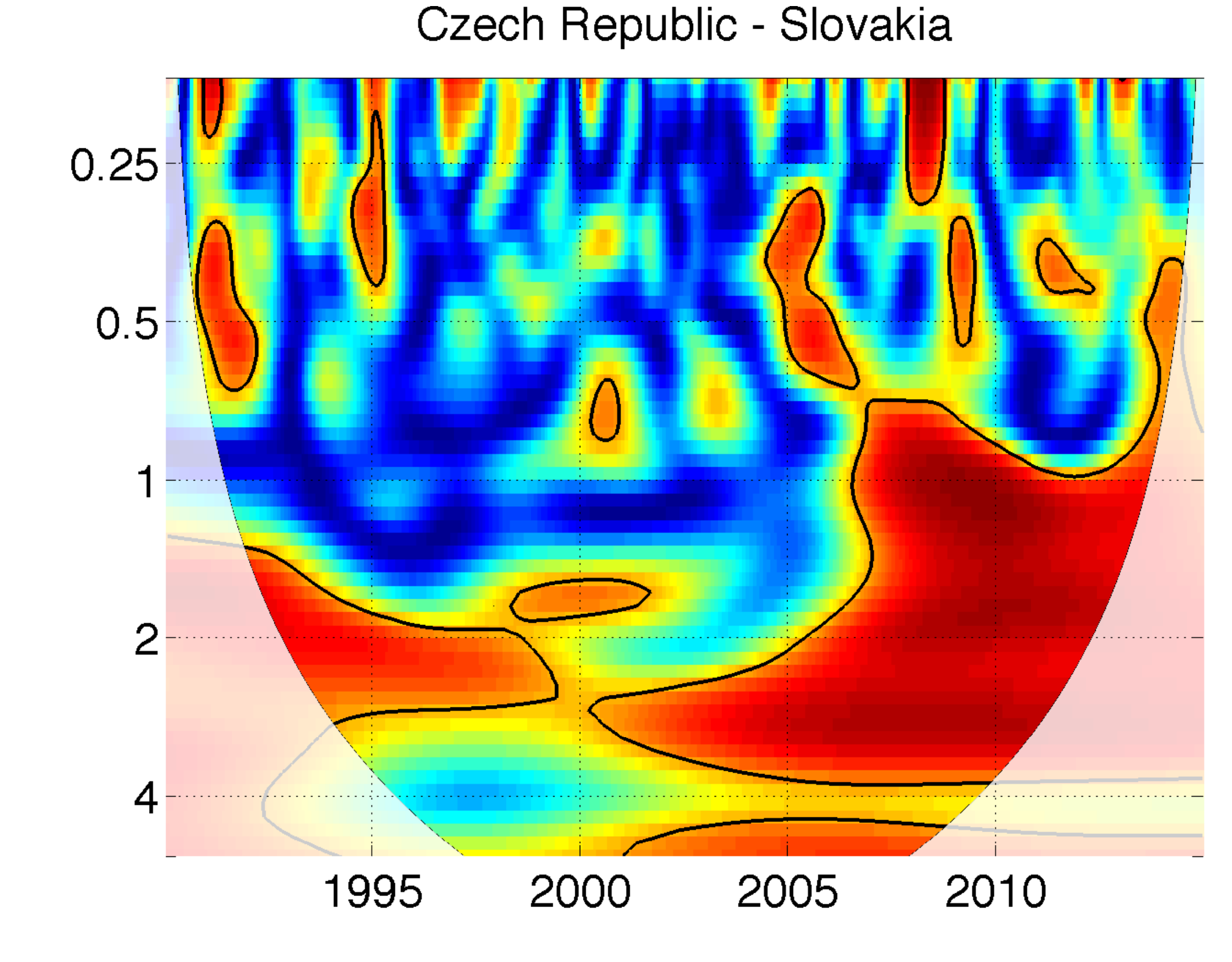}
  \end{minipage}
  \begin{minipage}{0.033\textwidth}
    \includegraphics[width=\textwidth]{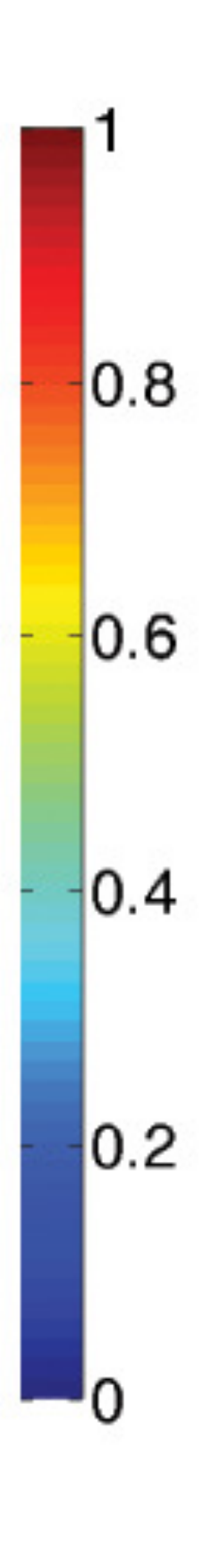}
  \end{minipage}\\
  \begin{minipage}{0.31\textwidth}
    \includegraphics[width=\textwidth]{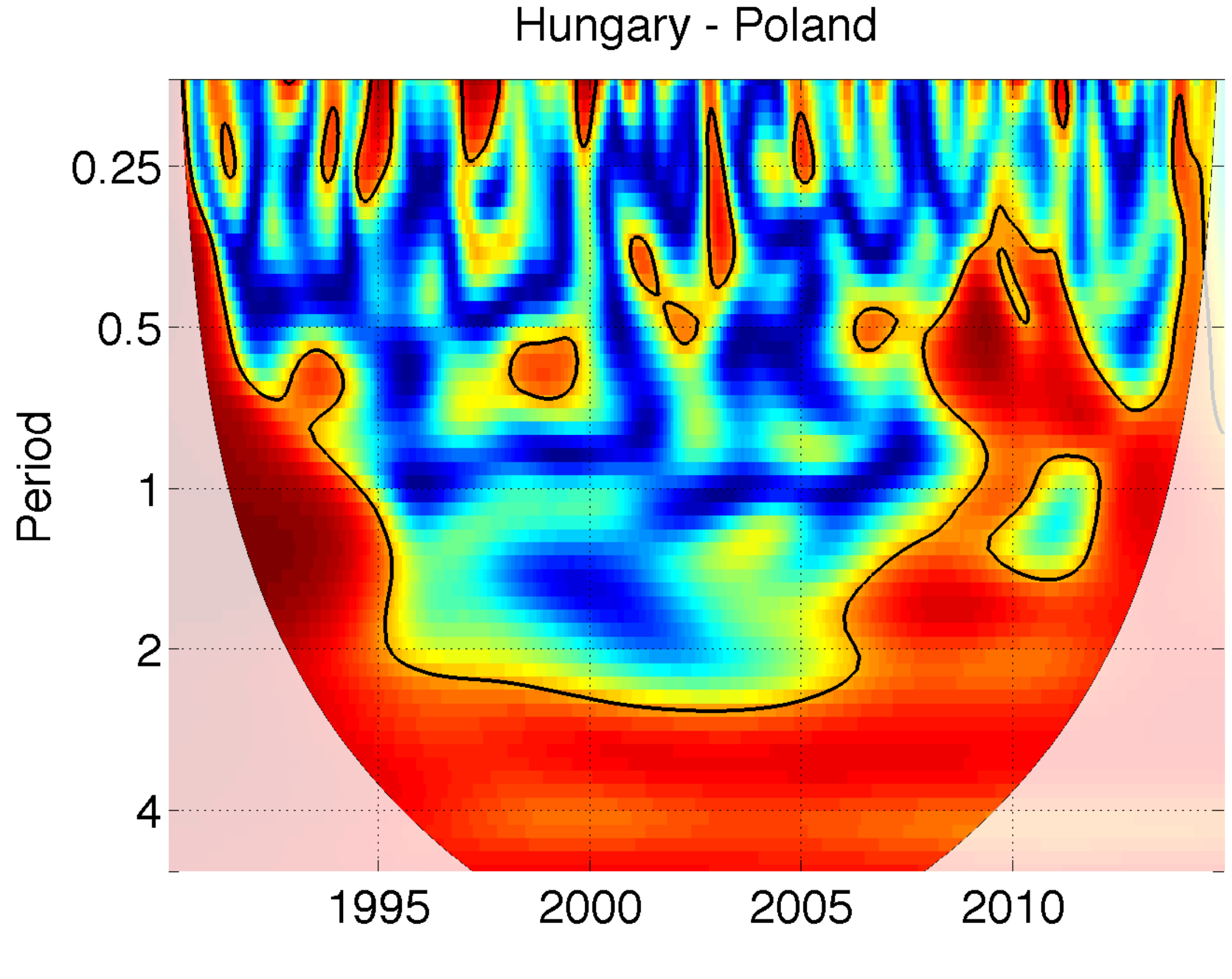}
  \end{minipage}
  \begin{minipage}{0.31\textwidth}
    \includegraphics[width=\textwidth]{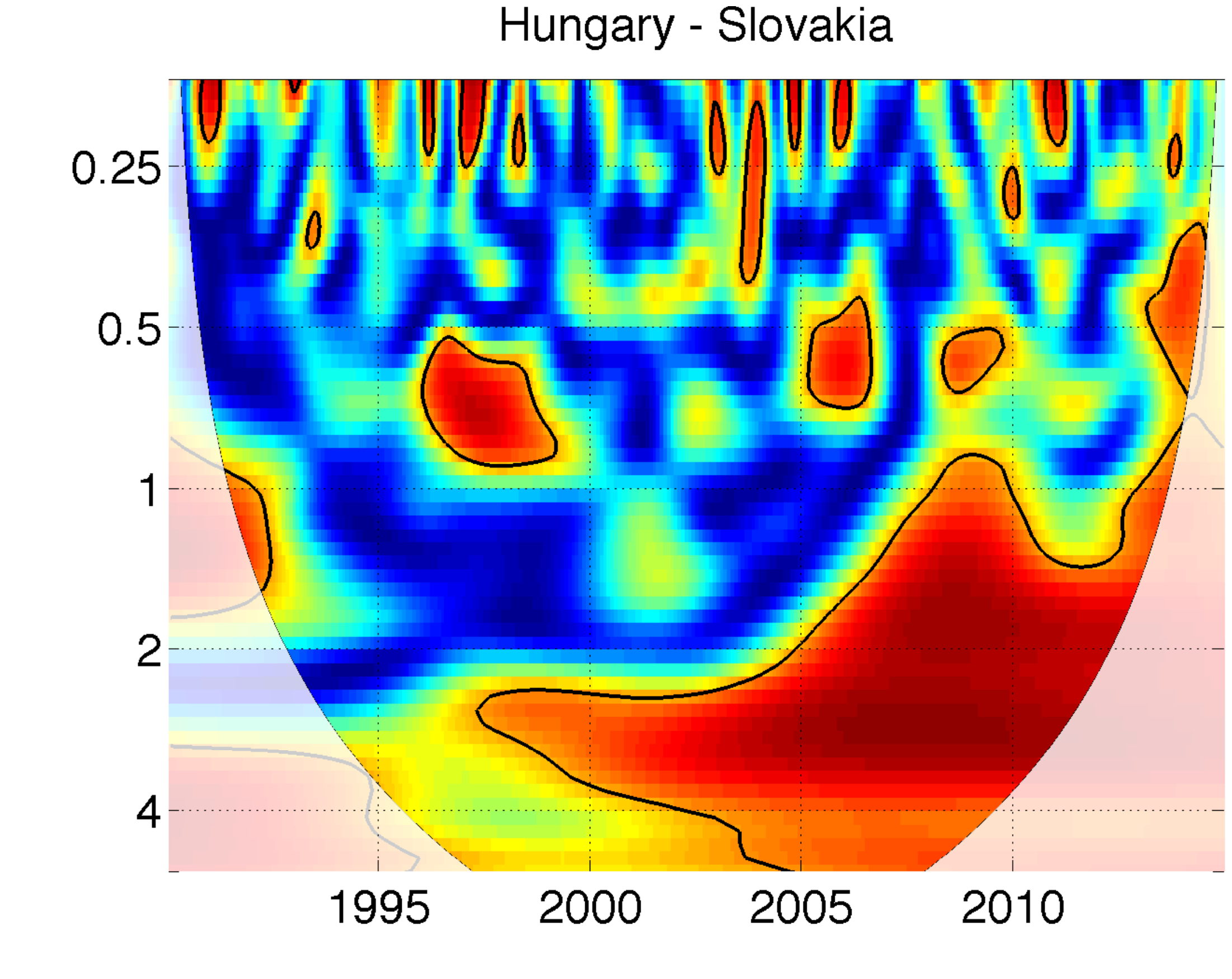}
  \end{minipage}
  \begin{minipage}{0.31\textwidth}
    \includegraphics[width=\textwidth]{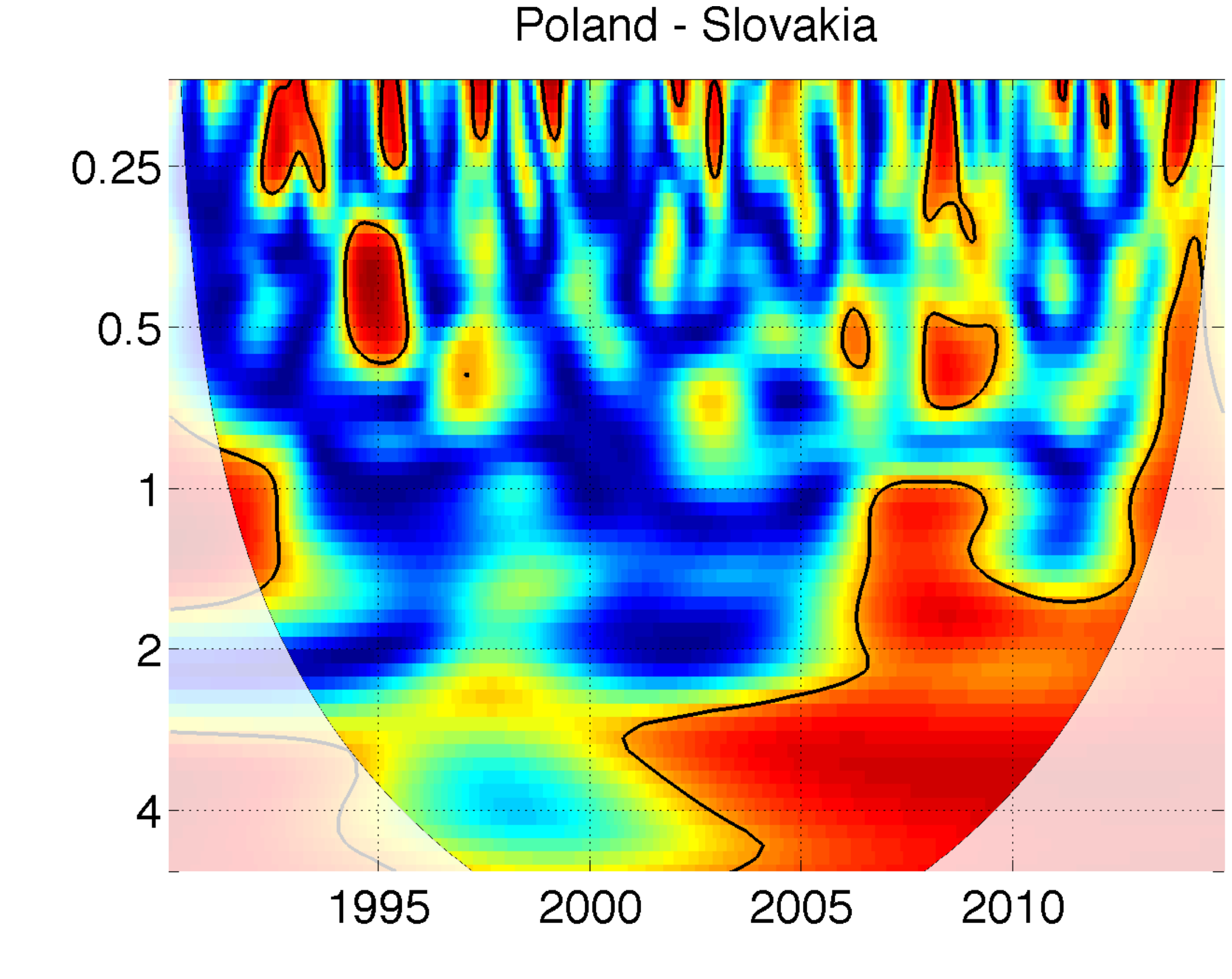}
  \end{minipage}
  \begin{minipage}{0.033\textwidth}
    \includegraphics[width=\textwidth]{bar}
  \end{minipage}
  \caption[Wavelet coherence within the Visegrad Four]{Wavelet coherence within the Visegrad Four. The significance level of 5\% against the red noise is contoured by the solid black line. The shaded area is the cone of influence.}
  \label{fig:withinVisegrad}
\end{figure}

At first, we analyze the business cycles of the V4 on an intra-group basis to disclose similarities in their pairwise co-movements and the development of particular relationships within the group.
The V4's cooperation began in the early 90s, and the countries share higher coherence at the beginning of the transition for 1-2 year period\footnote{In the text, we use notions as "1-2 year period", which correspond business cycles at this frequency (period length), thus 1-2 year business cycle.} during 3-4 years, see Fig.~\ref{fig:withinVisegrad}. Another common feature among the V4 countries is a weak relationship of all pairs at short-term business cycles, from 2 months to 1 year during whole 25 years. Only Hungary and Poland co-move significantly around 2010 for periods shorter than one year. The important result is that all pairs show a high degree of synchronization over the 2-4 year period beginning around 1998 for Hungary with Slovakia, 1999 for the Czech Republic with both Poland and Hungary, and 2000 for Poland with Slovakia. In addition, the Czech Republic and Slovakia have been synchronized over a 2-year business cycle period through the whole sample period with a small decrease around 2000. Hungary and Poland show a high degree of synchronization at all business cycle frequencies, 2-5 years. Their co-movement covers the largest part of the frequency spectrum at the beginning of the transition period and during the last 5 years.
% \vspace{-1em}

%The fact that at the beginning of transition countries shared high common volatility of business cycles across CEE countries \citep{Jagric:2002aa} is not shown in our results of co-movement.
We find a short time of higher synchronization for the first 3-5 years in 1-2 year periods. However, we see that all countries have almost zero co-movement around 1995, except for Hungary and Poland. This low degree of similarity may be caused by Slovakia's cold-shouldered participation in the political discussions during 1993-1997 that translated into the business cycles with a delay. Another possible explanation relevant for all countries can be that after a few years of formally intensive cooperation the monetary and fiscal policies started diverging. For example, during the late 1990s, the Czech Republic went through difficult stabilization years \citep{Antal:2008aa}. These diverging economic situations might cause some asynchrony in business cycle behavior over both the short- and long-terms. This long-term low synchronization may also come from the low level of convergence of other macroeconomic variables \citep{Kutan:2004aa}.

\begin{figure}[ht]
\centering
  \begin{minipage}{0.45\textwidth}
    \includegraphics[width=\textwidth]{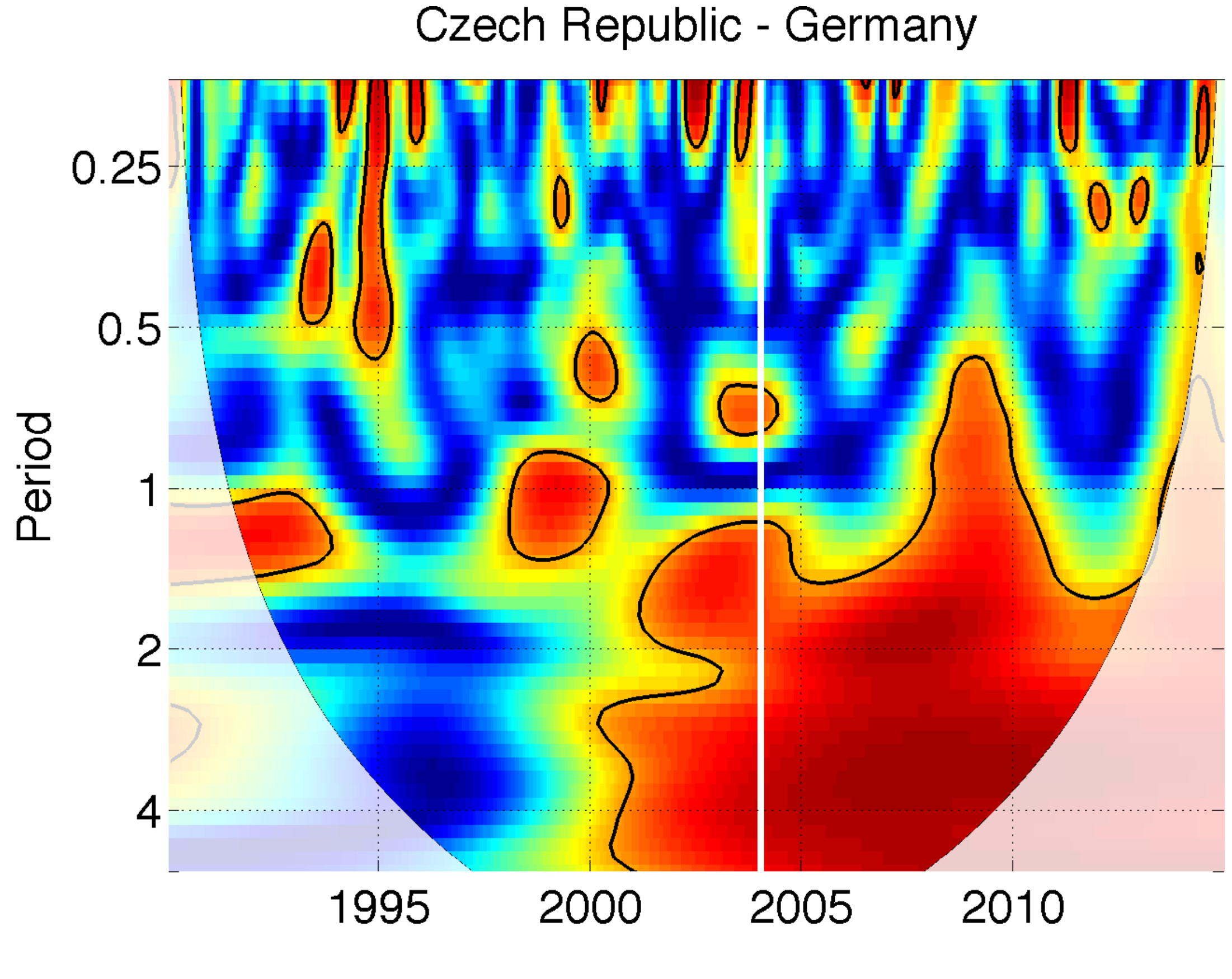}
  \end{minipage}
  \begin{minipage}{0.45\textwidth}
    \includegraphics[width=\textwidth]{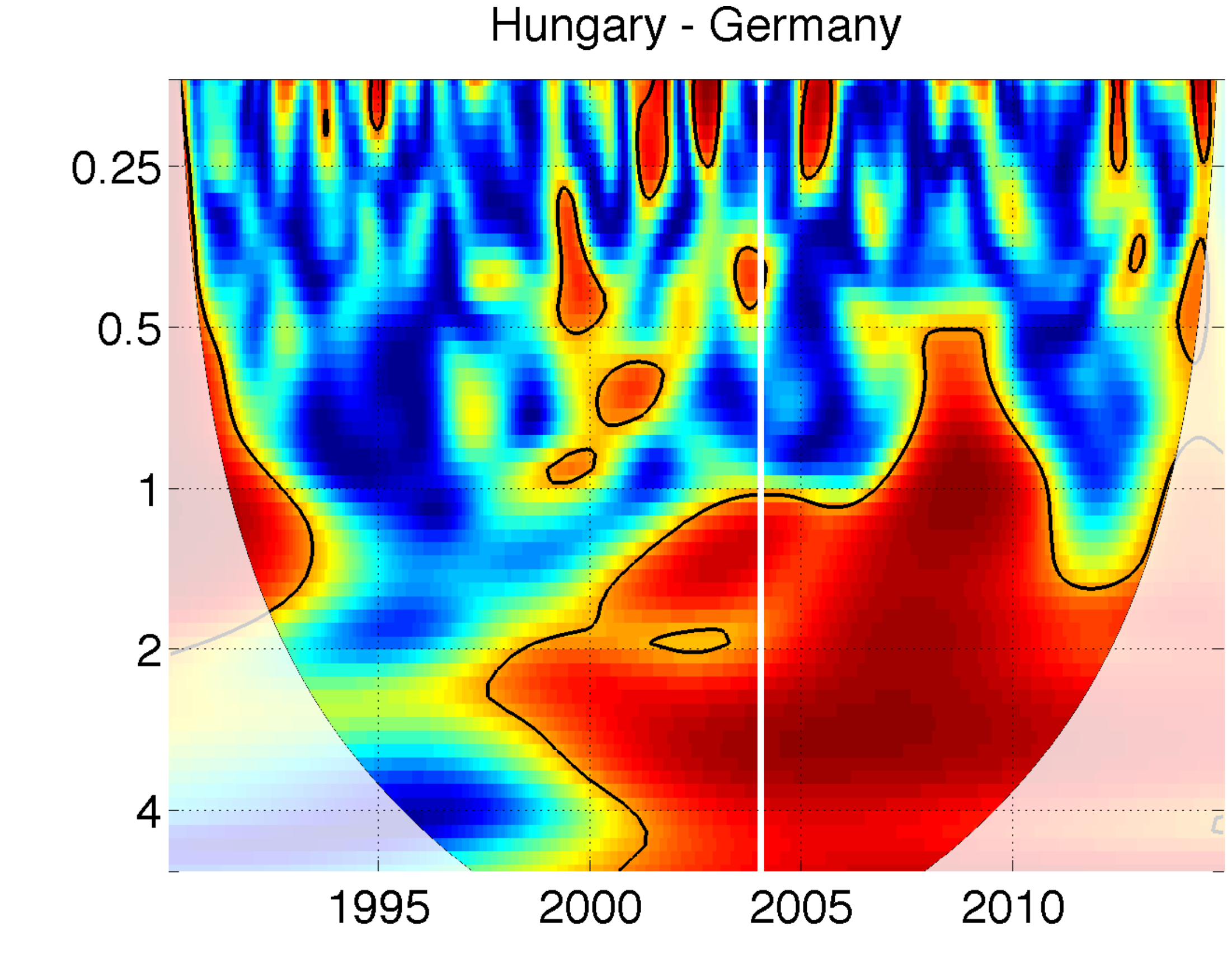}
  \end{minipage}
  \begin{minipage}{0.047\textwidth}
    \includegraphics[width=\textwidth]{bar}
  \end{minipage}\\
  \begin{minipage}{0.45\textwidth}
    \includegraphics[width=\textwidth]{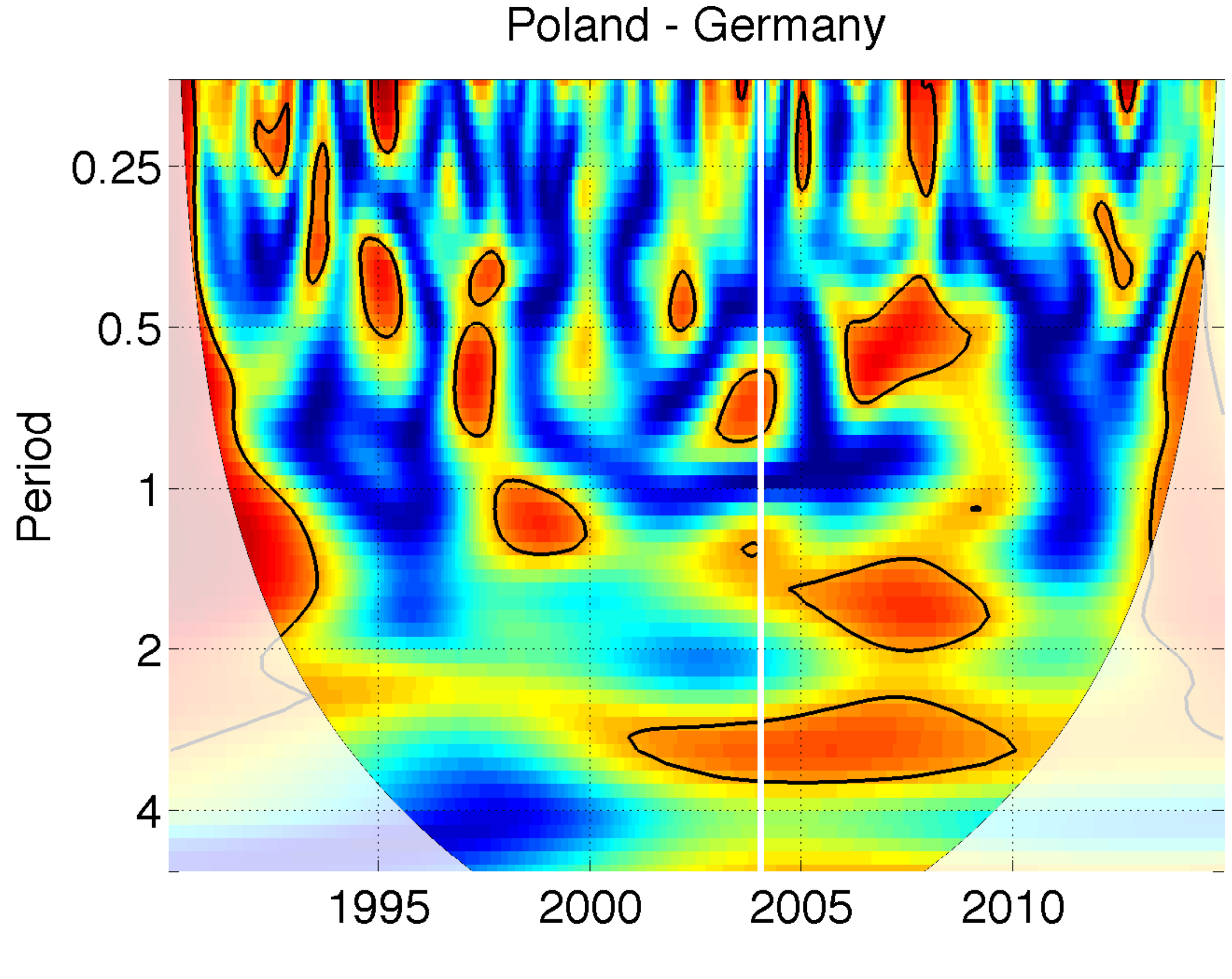}
  \end{minipage}
  \begin{minipage}{0.45\textwidth}
    \includegraphics[width=\textwidth]{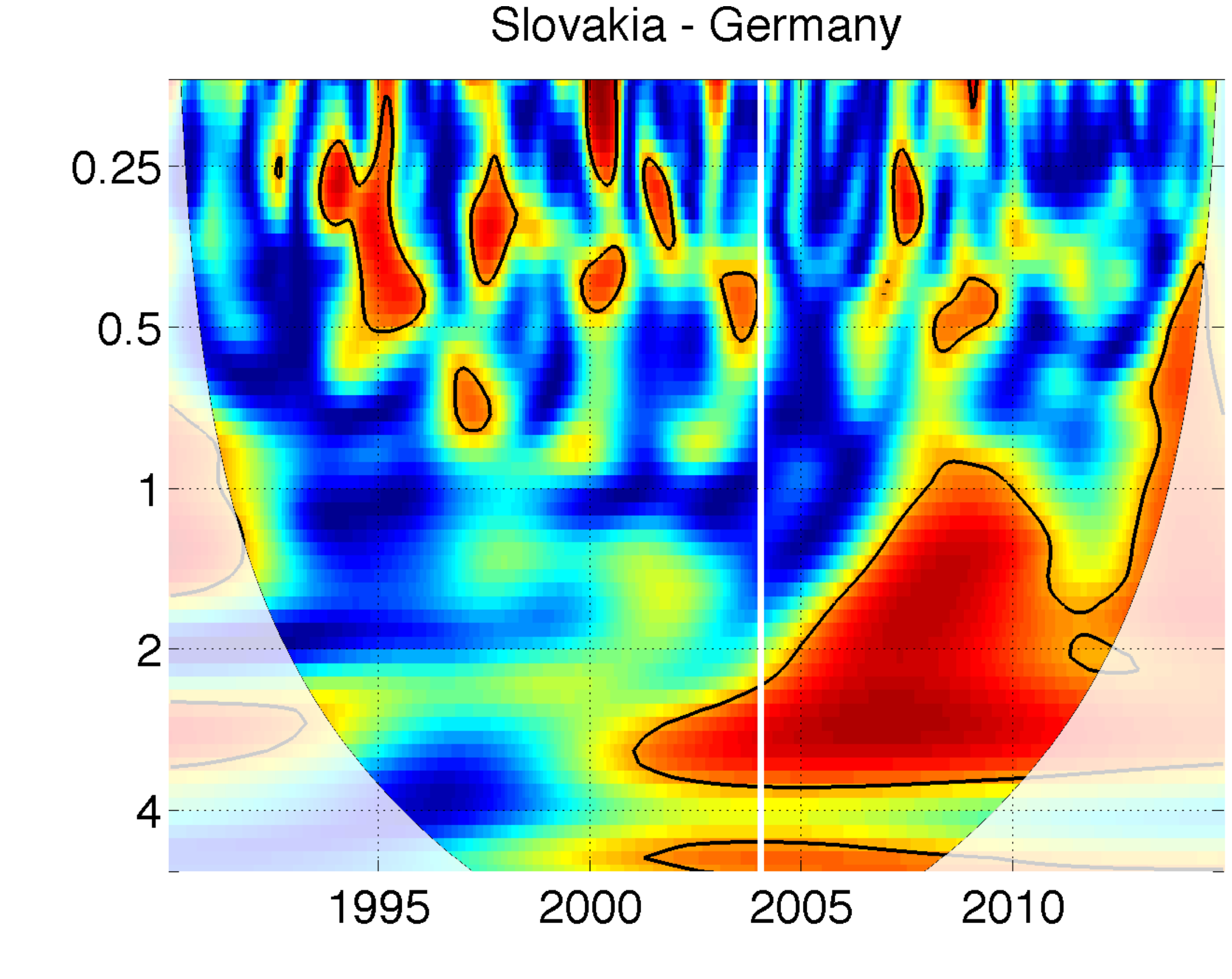}
  \end{minipage}
  \begin{minipage}{0.047\textwidth}
    \includegraphics[width=\textwidth]{bar}
  \end{minipage}
  \caption[Wavelet coherences of the Visegrad Four and Germany]{Wavelet coherences of the Visegrad Four and Germany. The vertical solid white line indicates 2004 -- the year of the enlargement of the EU. The significance level of 5\% against the red noise is contoured by the solid black line. The shaded area is the cone of influence.}
  \label{fig:visegradGermany}
\end{figure}

Further, we are interested in the co-movement of each country of the V4 within the framework of the European Union. In the pairwise analysis, we take Germany as a representative of the EU. Germany is often used as a reference country \citep{Fidrmuc:2006aa} for the EU. Moreover, a great portion of exports of V4 goes through Germany; meaning it is near the Visegrad region, which also plays a role. We plot the relationships between the V4 and Germany in Fig.~\ref{fig:visegradGermany}.

We observe the strongest co-movement of Germany with two countries: the Czech Republic and Hungary. The large area of high coherence for both begins around 2000, for a 1-5 year period. Hungary-Germany high coherence starts slightly before 2000 for a 2-year period but from 2000 it continues for a 1-5 year period onward. These findings are in line with the results of \citet{Aguiar-Conraria:2011aa}. The synchronization of Slovakia with Germany shows an interesting pattern. The coherence increases gradually from 2000 and spreads from 2-4 years to 1-4 years period around 2008, which is important because that is precisely when Slovakia adopted the Euro, on January 1, 2009. Slovakia may be also considered as an example where the degree of synchronization increases after accession to the EU and EMU, which is consistent with the theory of endogeneity of optimum currency areas. This high coherence may be a reaction to the global financial crisis; however, it may support higher synchronization with the EU as the shock would spill over the euro area to countries, such as those of V4. For all the V4, we see an increase of coherence after beginning their preparations for EU accession, which was shortly before 2000, this support the finding of \citet{Kolasa:2013aa} who reports a substantial convergence with the Eastern enlargement of the EU. The high degree of synchronization of Hungary supports also the previous results of \citet{Fidrmuc:2006aa}, for example. However, this same support is not true for Poland because we observe the weakest relationship between Poland and Germany, although we may find a few isolated islands of higher coherence, which are unimportant in comparison with other countries.
\begin{figure}[ht]
\centering
  \begin{minipage}{0.47\textwidth}
    \includegraphics[width=\textwidth]{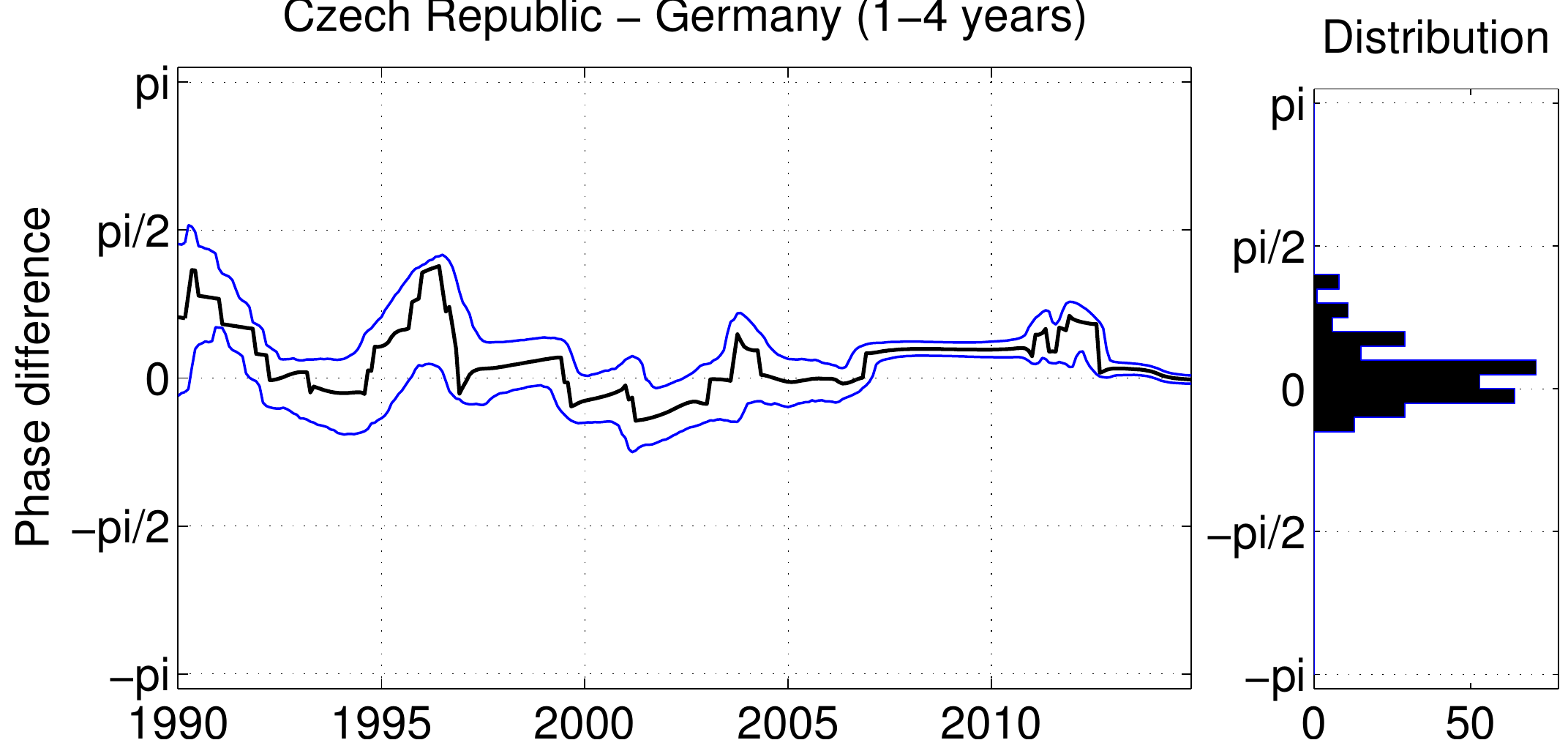}
  \end{minipage}
  \begin{minipage}{0.47\textwidth}
    \includegraphics[width=\textwidth]{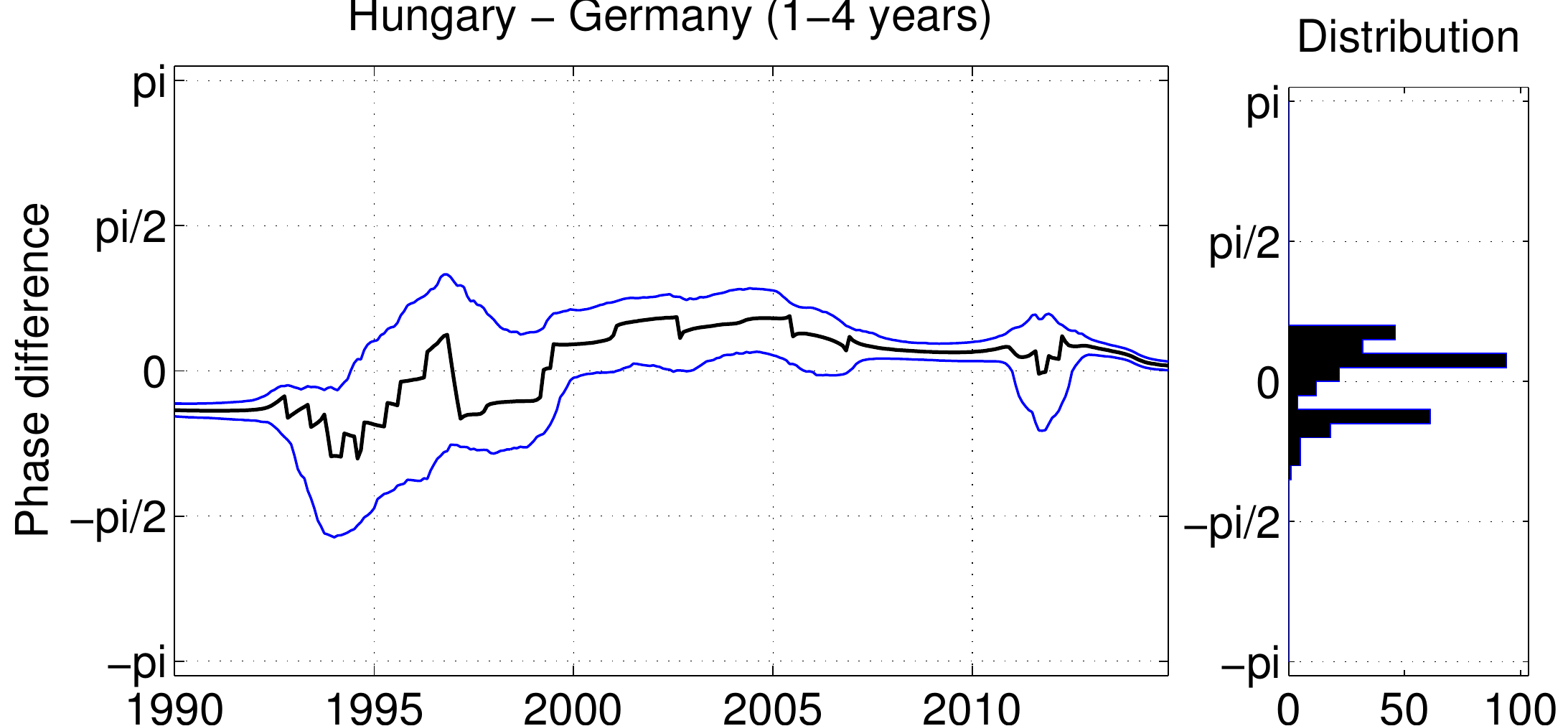}
  \end{minipage}\\[0.2em]
  \begin{minipage}{0.47\textwidth}
    \includegraphics[width=\textwidth]{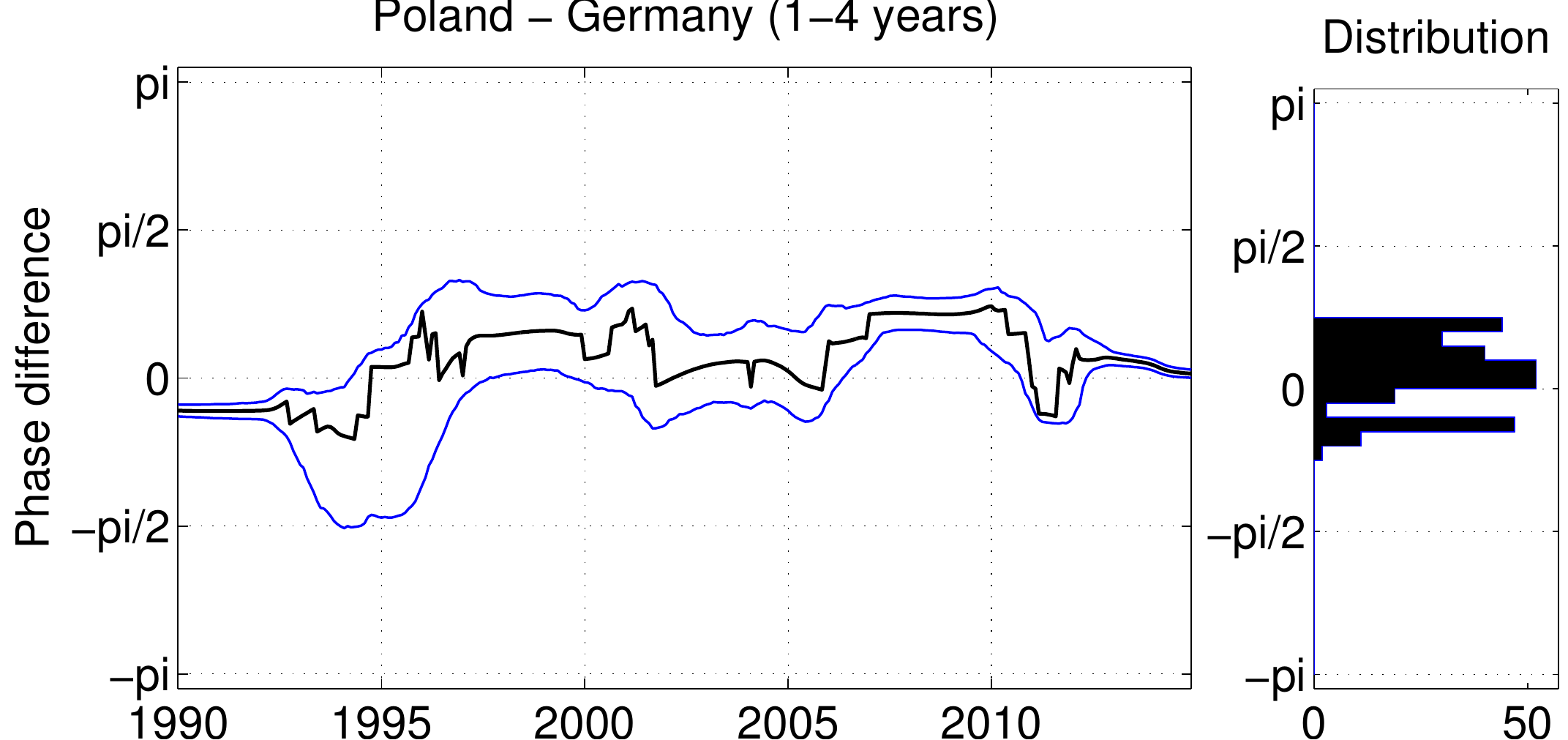}
  \end{minipage}
  \begin{minipage}{0.47\textwidth}
    \includegraphics[width=\textwidth]{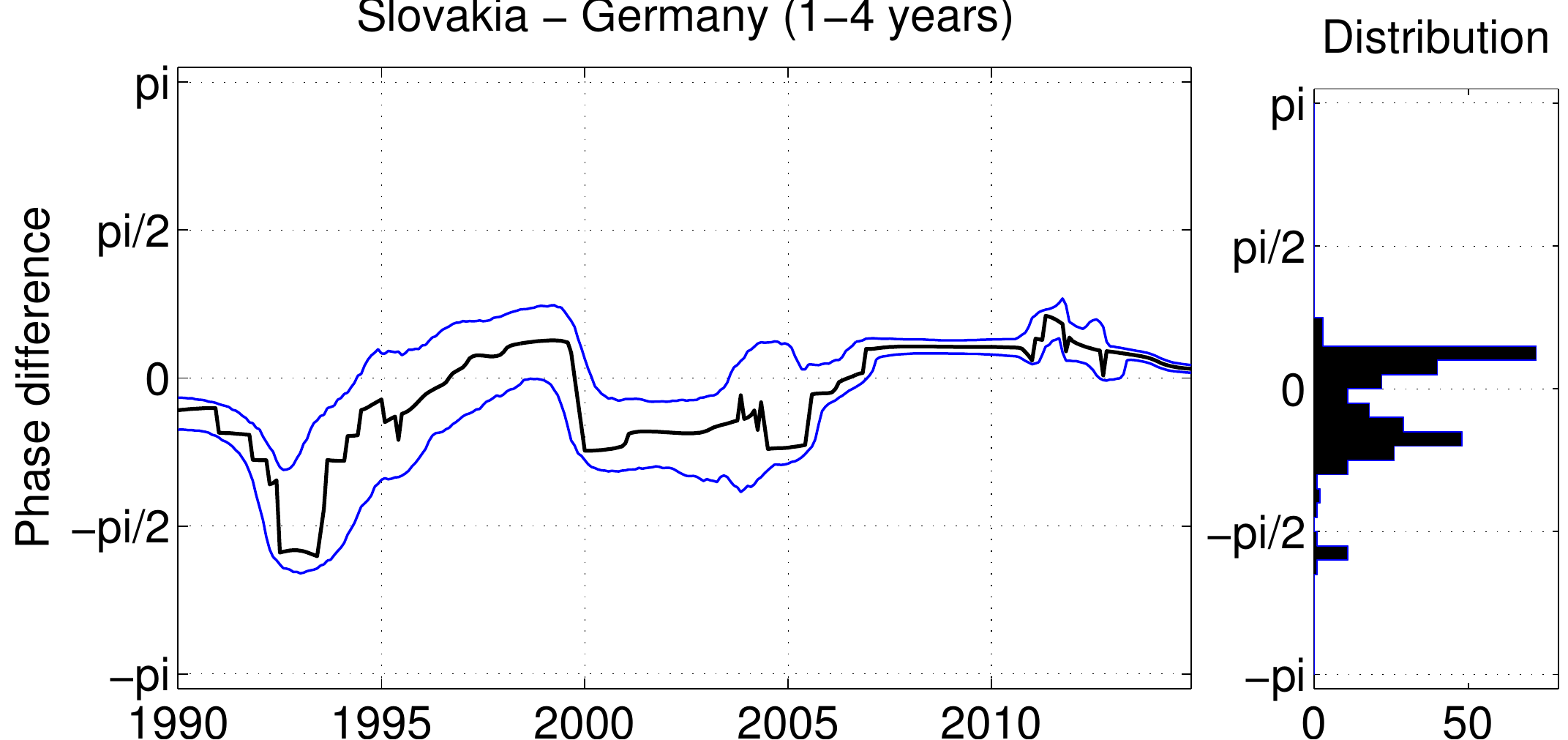}
  \end{minipage}\\[0.15em]
  \caption[Phase differences]{Phase differences. The black solid line is the true phase difference of two time series. The blue solid line is the 95\% bootstrapped confidence interval. For each phase difference, its distribution is provided.}
  \label{fig:phases}
\end{figure}

Additionally, we provide the analysis of phase differences between each V4 nation and Germany. The phase difference for the 1-4 year period fluctuates around zero most of the time, which corresponds to changing the lead position between two countries that have a positive relationship -- in phase. We present confidence intervals of these phase differences. In the most cases, as the phases fluctuate the confidence intervals overlap zero; thus, we cannot determine for certain which country is in the lead position. Further, the first and last five years of data are affected by edge effects of zero-padding; thus, we do not have complete information here. Following \citet{Aguiar-Conraria:2014aa}, we rely on the significance of coherence to know whether the phase difference is also significant. The coherence is found high and mostly significant for all countries between 2000 and 2010. Fig.~\ref{fig:phases} shows that the phase difference is significant in 2006-2010 period and it belongs to the $[0,\pi]$ interval, which also shows that the V4 business cycles lead the cycles of Germany over the long-term period during the 2006-2010 period. One possible explanation for this finding can be that recessions or rebounds in the productions of the V4 contries happen sooner compared to Germany. For example, in case of Slovakia, we have the significant leading position of Germany during 2000-2004. Showing that the bootstrapped confidence intervals, we demonstrate that this technique is relevant and desired for conclusion about phase differences between time series. %Even if the phase difference clearly belongs to some interval, we need some level of confidence to make a conclusion.

\subsection{Multivariate co-movement}

Until now, we have used the bivariate analysis, and we have omitted the assessment of synchronization of more than two time series. In this part, we investigate the multivariate relationship of countries in the EU. The proposed measure of cohesion with time-varying weights allows us to assess the co-movement within the groups of countries. %This measure captures the information in the time-frequency plane; we provide the results in the form of a heat-map as in the previous section.
In contrast to the coherence, the cohesion may be negative, which means it can capture a counter-cyclical co-movement among the time series. We employ the time series of Gross Domestic Product (GDP) in current prices and GDP at purchasing power parity (PPP) per capita to weight the economic activity. % of a country by its size.
The choice of GDP as weights is particularly to quantify the size of countries' economies. Compared to the fixed weights of \citet{Rua:2012aa}, the time-varying weights take into account different developments of countries.\footnote{In the multivariate analysis, we analyze a period of 17 years spanning from 1997M1-2014M3 due to the lack of data for some countries.}

% \subsubsection{Example of different weights in the multivariate case}
\subsubsection{Does the size of economies affect business cycle cohesion?}

A broader picture about the business cycle synchronization in the region of the V4 countries and Germany can be studied using the cohesion measure. Since the beginning of the transition, we have observed a significant growth of nominal GDPs of the Visegrad countries, see Table~\ref{tab:gdp1}. % during the studied time span.
Slovakia's nominal GDP grew by almost 300\%, the nominal GDPs of Czech Republic, Hungary and Poland have more than doubled by that time, whereas the nominal GDP of Germany increased only by 52\%. The time-varying weights allow us take into account  gradual changes of proportions between economies and reflect this development in their business cycle synchronization. The evolving size of economies can project and emphasize possible convergence or divergence among economies. Further, we look at the GDPs at power purchasing parity per capita where, at smaller extent, the same happen. GDPs at PPP per capita more than doubled for the Visegrad countries and for Germany it grew by 75\%, almost doubled.
\begin{table}[!ht]
\centering
\caption{Change of GDPs}
\label{tab:gdp1}       % Give a unique label
\begin{tabular}{lcccccc}
\hline\noalign{\smallskip}
1997 vs. 2014 & Germany & Czech Rep. & Hungary & Poland & Slovakia \\
\noalign{\smallskip}\hline\noalign{\smallskip}
$\Delta$ GDP level (in \%) & 52  & 169 & 133 & 196 & 290 \\
$\Delta$ GDP at PPP (in \%) & 75 & 102 & 110 & 165 & 151 \\
\noalign{\smallskip}\hline
\end{tabular}
\captionsetup{width=0.9\textwidth}
\caption*{\textit{Note:} The values show by how many per cents the GDPs of given countries have grown between 1997 and 2014.}
\end{table}
\vspace{-1.5em}

This motivates our new approach to show clearly that the nature of weighting in such measure is crucial. The cohesion puts a weight to each pair in the multivariate analysis with respect to all variables. For instance, when considering Germany and Poland within the group weighted by nominal GDP, their pair has the largest effect on cohesion. Although, their pairwise co-movement is weak and thus, it lowers the multivariate cohesion at many points with respect to weights. In contrary, from previous analysis we know that the Czech Republic and Hungary strongly co-move with Germany in the long-term frequencies, thus the effect on the cohesion can be larger because the presence of Germany in the pair.
\begin{figure}[!ht]
\centering
  \begin{minipage}{0.44\textwidth}
    \includegraphics[width=\textwidth]{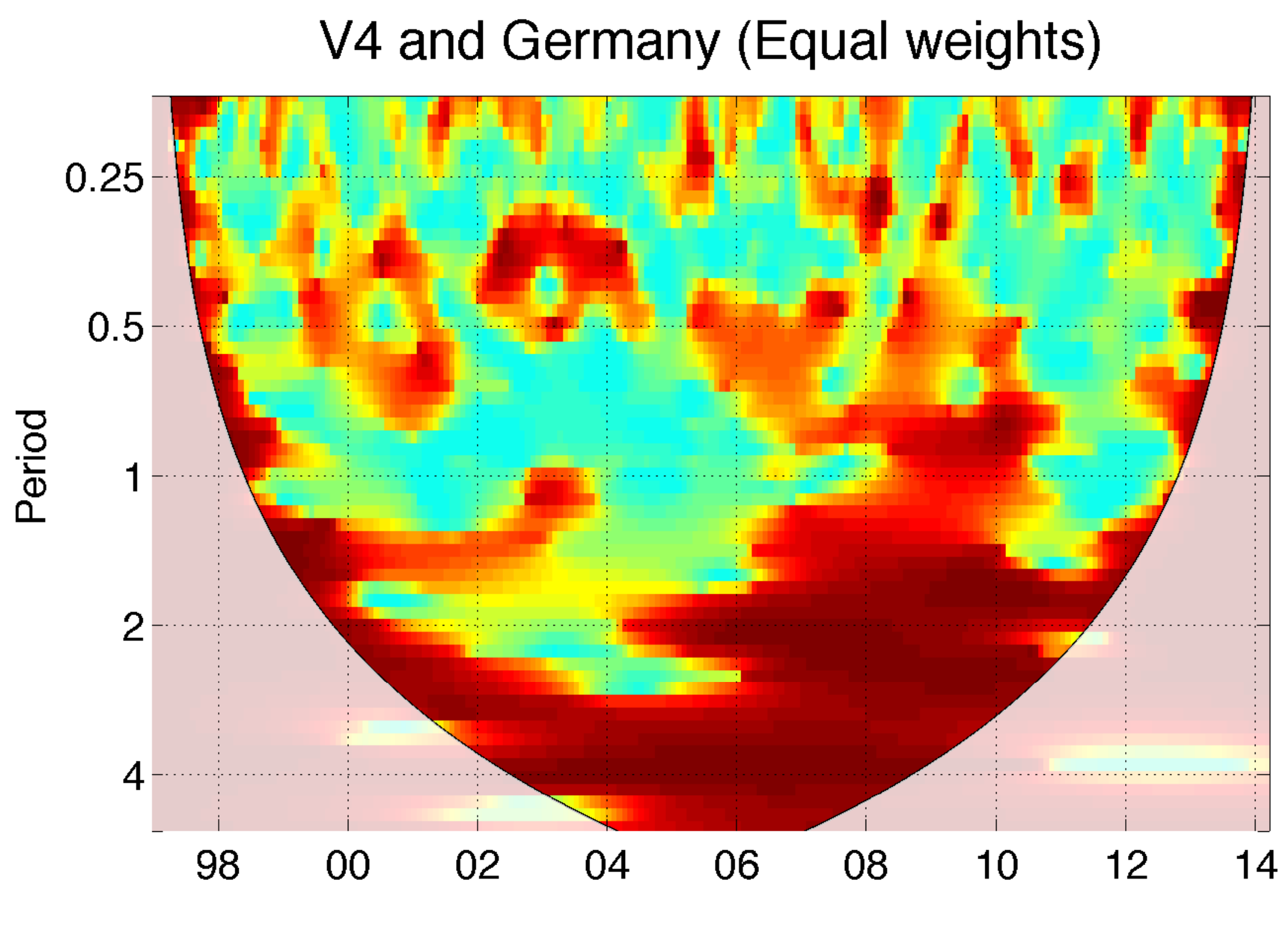}
  \end{minipage}
  \begin{minipage}{0.47\textwidth}
    \includegraphics[width=\textwidth]{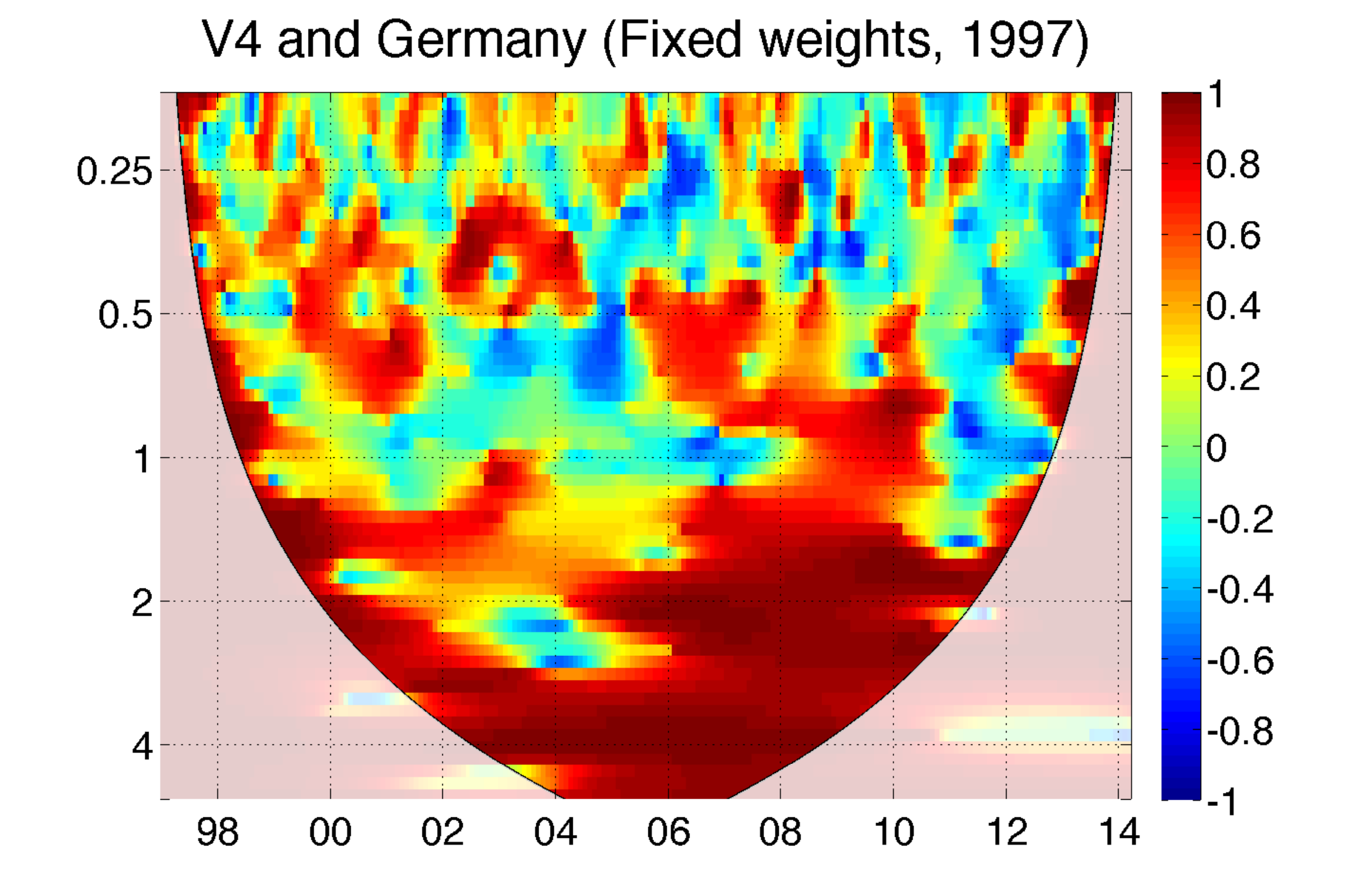}
  \end{minipage}\\
  \begin{minipage}{0.44\textwidth}
    \includegraphics[width=\textwidth]{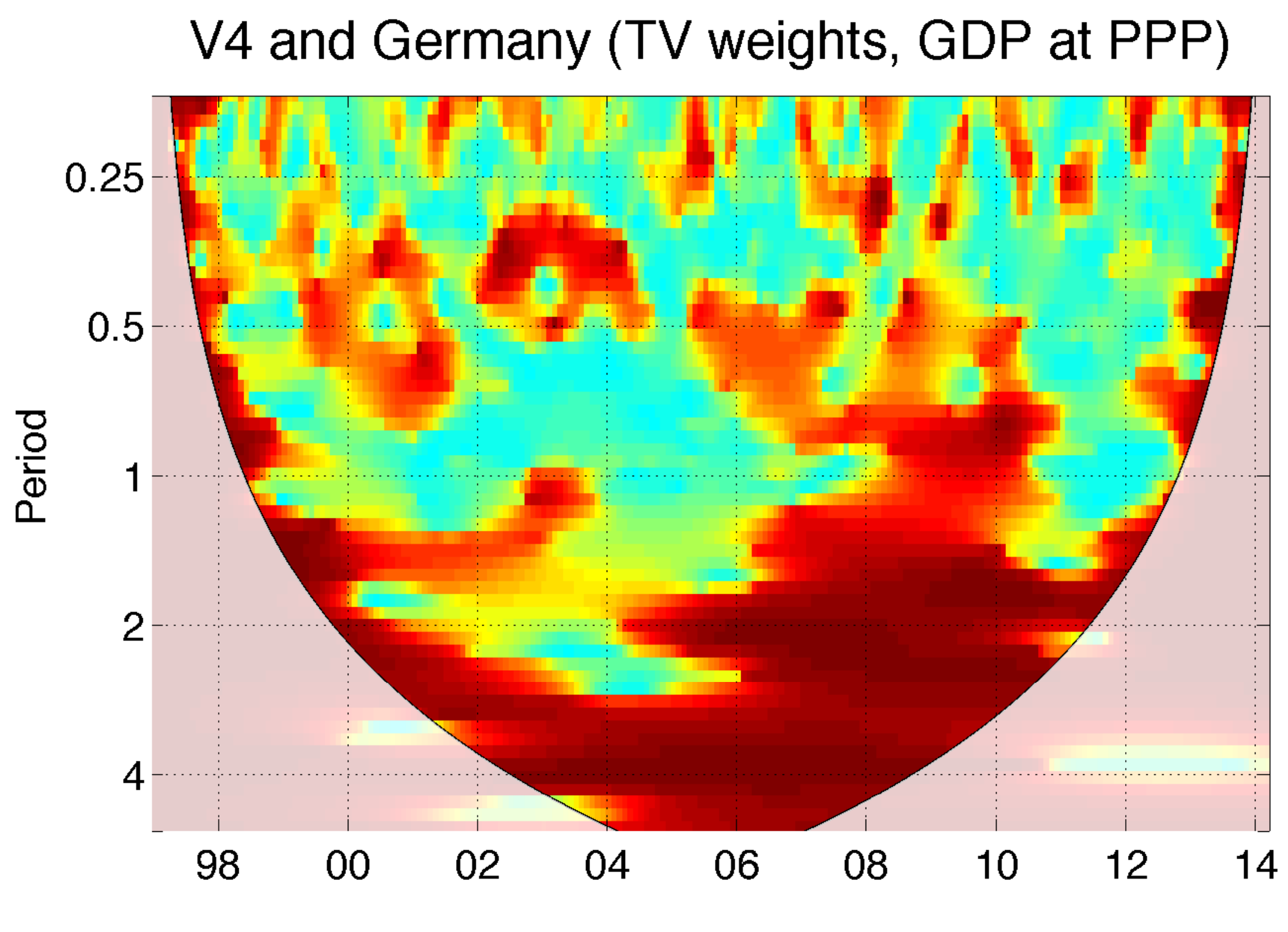}
  \end{minipage}
  \begin{minipage}{0.47\textwidth}
    \includegraphics[width=\textwidth]{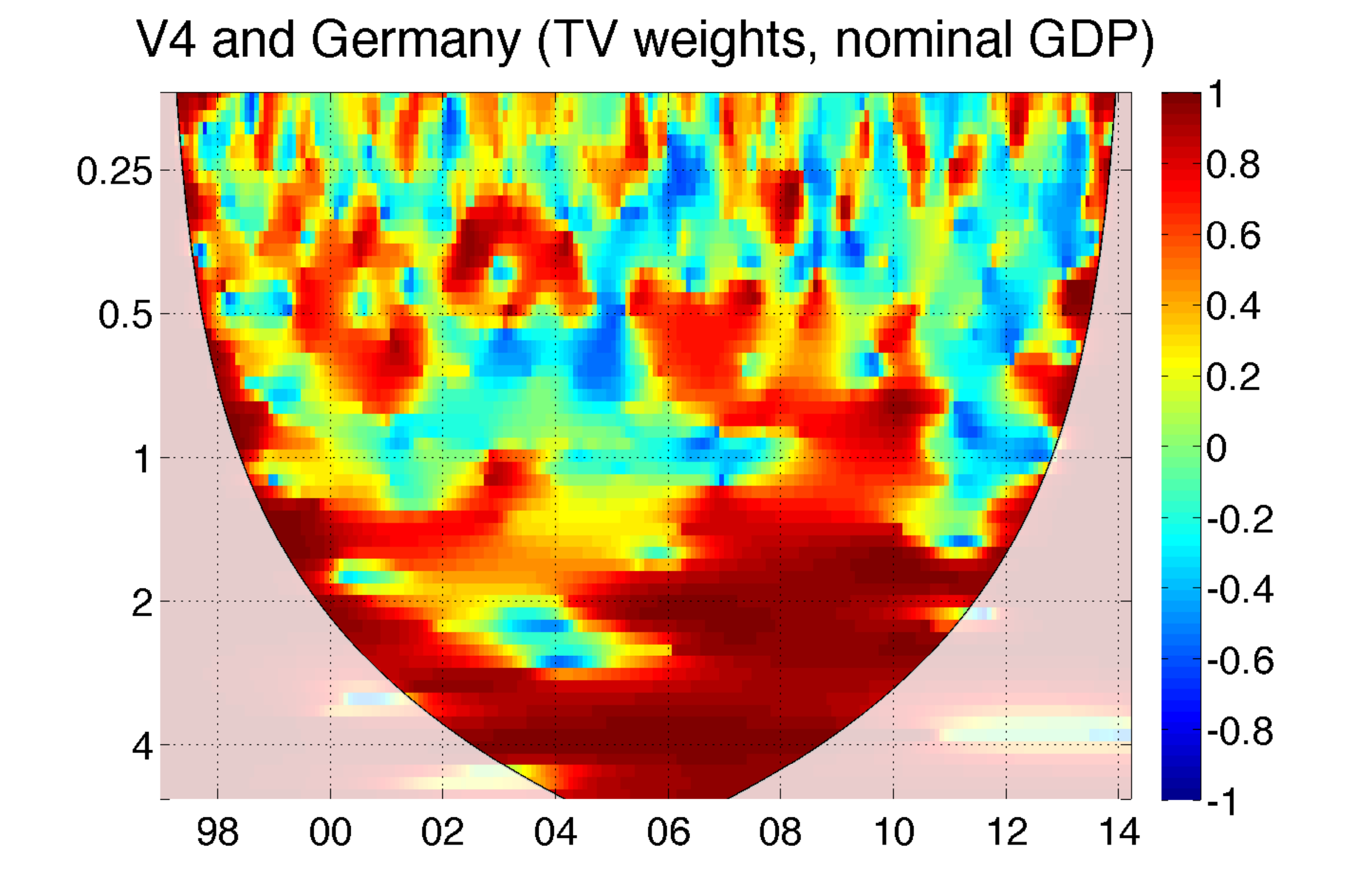}
  \end{minipage}
  \caption[Wavelet cohesion]{Wavelet cohesion of the Visegrad Four and Germany using different type of weights: equal (top-left), fixed (top-right), time-varying GDP at PPP (bottom-left), and time-varying GDP level (bottom-right).}
  \label{fig:cohWeights}
\end{figure}

\begin{figure}[!ht]
  \begin{minipage}{0.33\textwidth}
    \includegraphics[width=\textwidth]{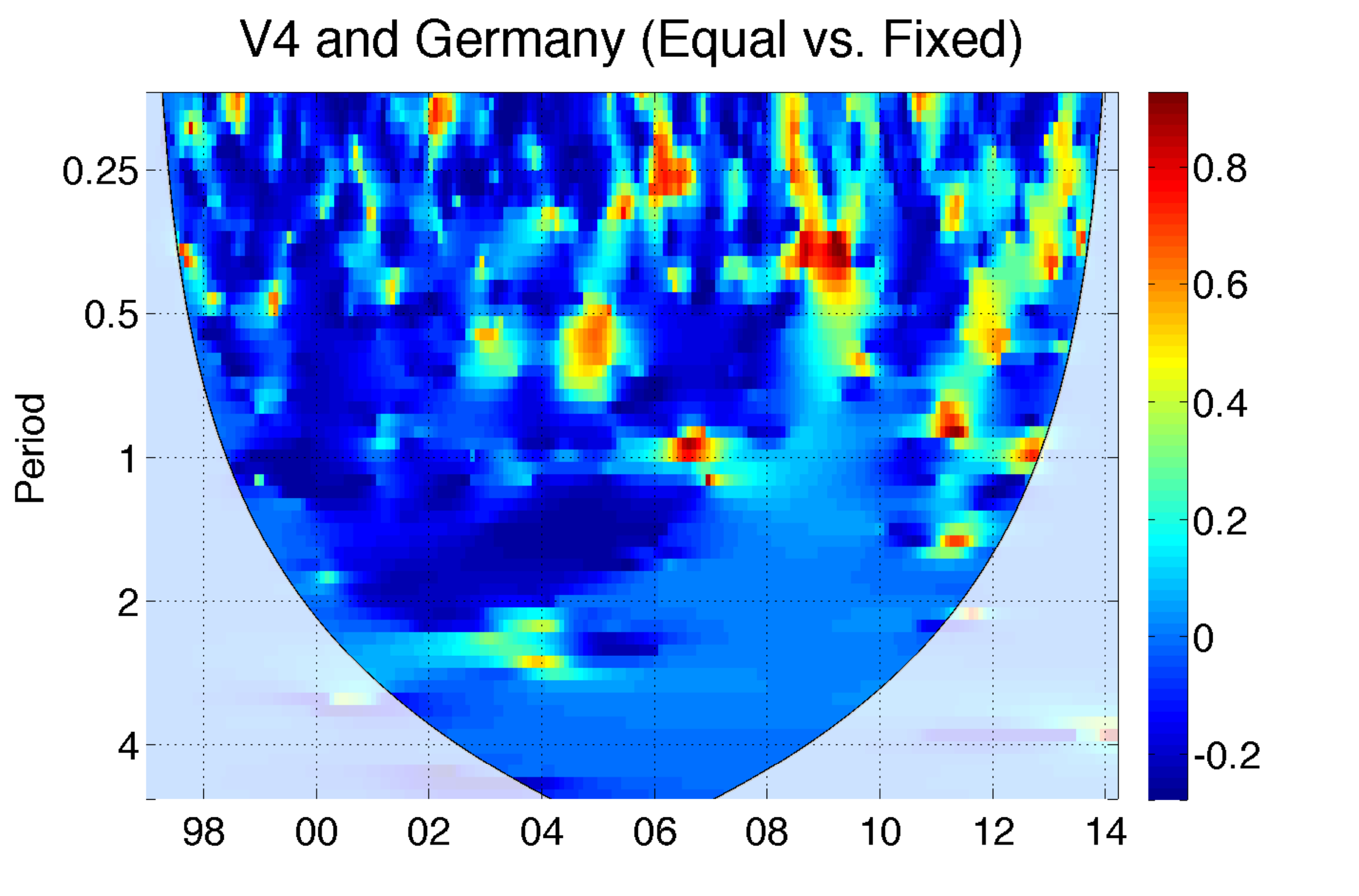}
  \end{minipage}
  \begin{minipage}{0.32\textwidth}
    \includegraphics[width=\textwidth]{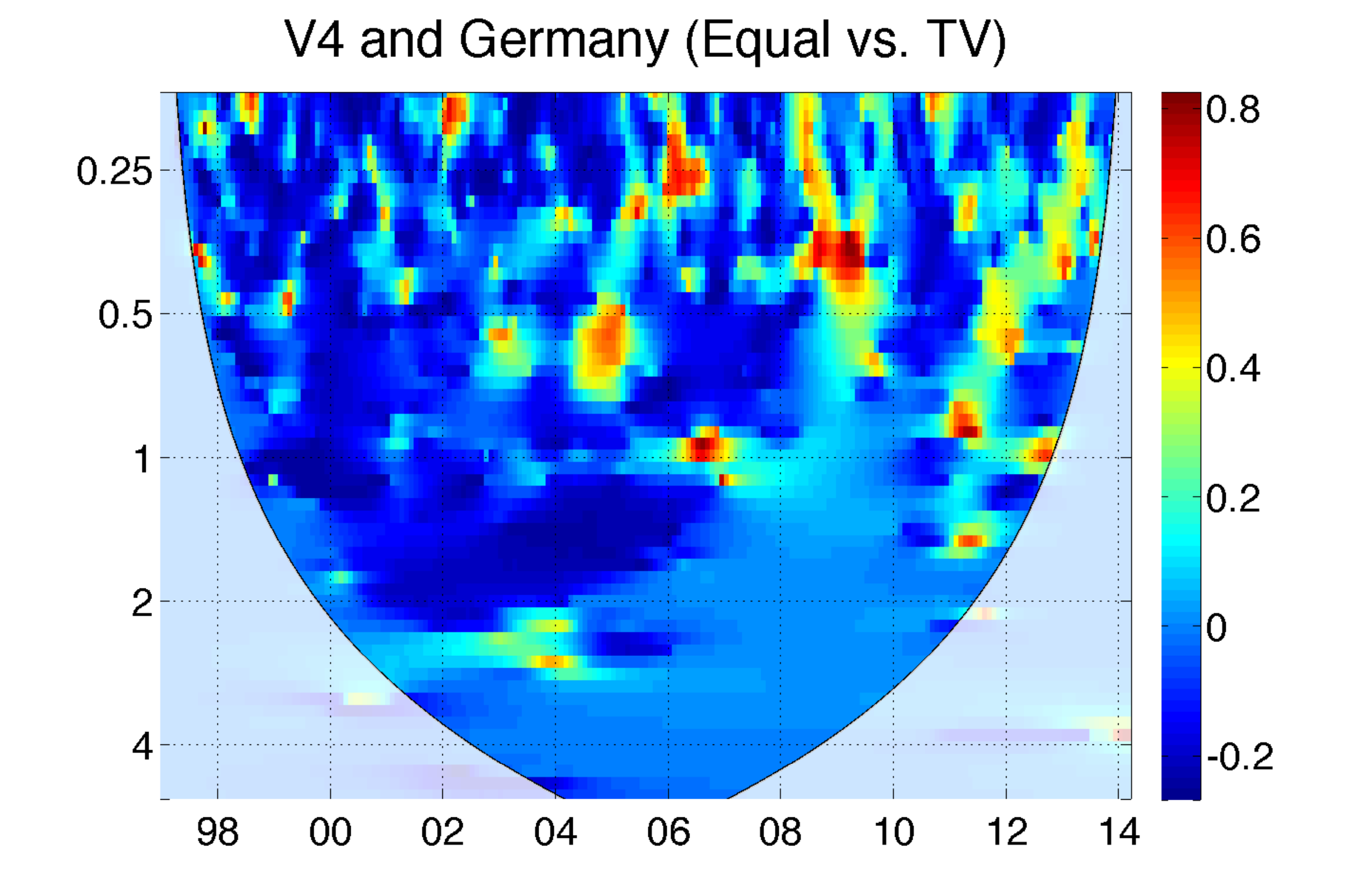}
  \end{minipage}
  \begin{minipage}{0.32\textwidth}
    \includegraphics[width=\textwidth]{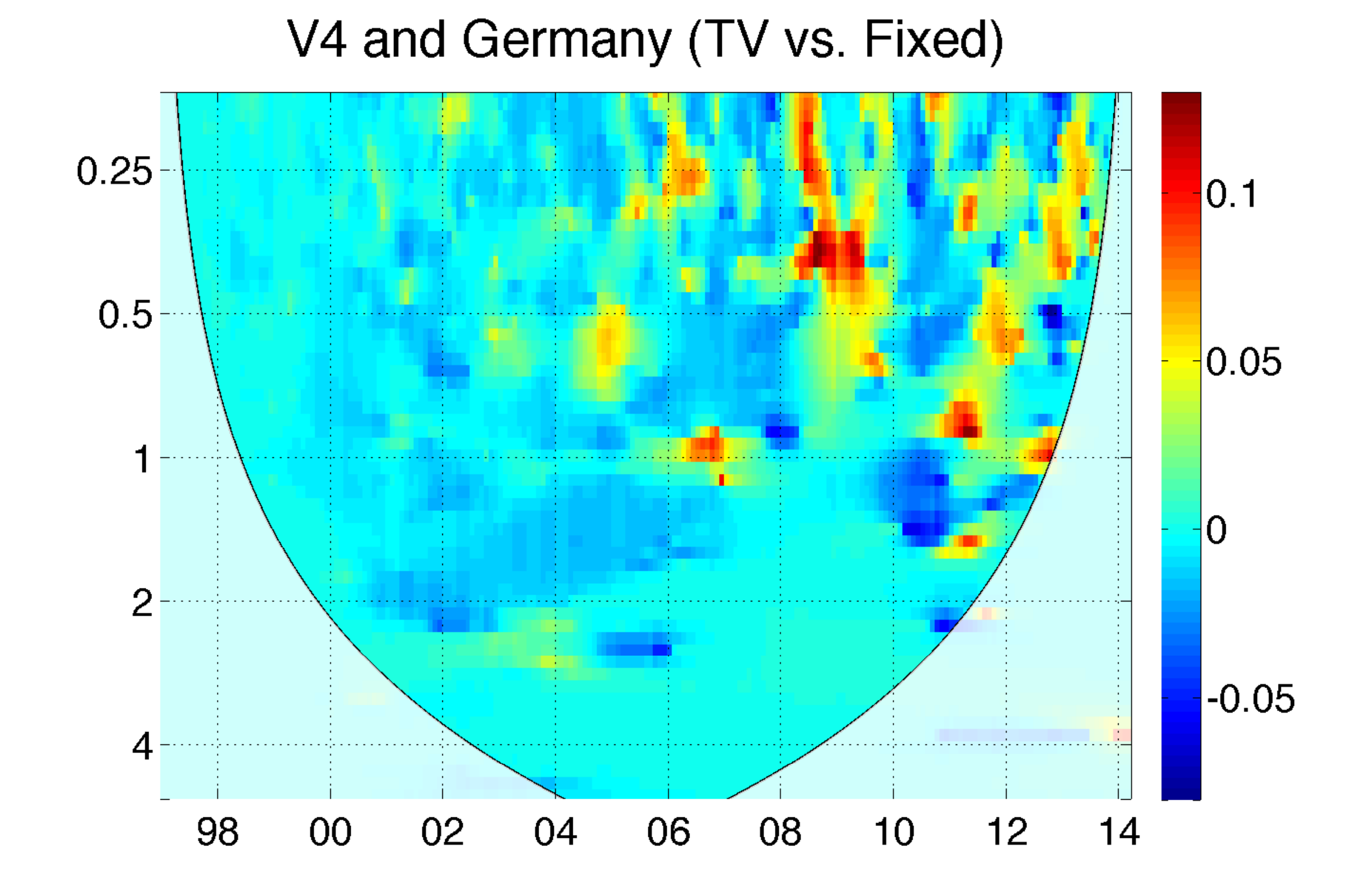}
  \end{minipage}
  \caption[Difference between weights]{Differences between cohesions from Figure~\ref{fig:cohWeights}: equal vs. fixed (left), equal vs. time-varying (center), and fixed vs. time-varying (right).}
  \label{fig:diffCohWeights}
\end{figure}

To illustrate the advantage of dynamic weights, we compare four cases of weighting: equal weights -- each country has equal size, fixed weight corresponding to one moment -- 1997Q1 (Eq.~\ref{eq:ruaCoh}), and then two cases of employing time-varying weights (Eq.~\ref{eq:myCoh}). Further, the distinction was made and we show the differences between cohesions that employing equal, fixed, and time-varying weights yields different results, Fig.~\ref{fig:cohWeights} and~\ref{fig:diffCohWeights}. %The dynamic multivariate relationship of the V4 and Germany looks very similar for all three types of weights, see Fig.~\ref{fig:cohWeights}

In Fig.~\ref{fig:cohWeights}, we observe high cohesion of 2-4 year business cycles of the Visegrad countries and Germany. Further, the cohesion is quite mild (around zero) at one-half to 1 year business cycle period. There appear several areas of positive cohesion at shorter periods, as well as, few small areas of negative cohesion distributed over periods from a quarter to 1 year. In the case when the nominal GDP is used, we observe that the negative areas of cohesion are not that sharp as when fixed weights were used. This refers that changes of proportions between economies lower counter-cyclicalities in cohesion of the Visegrad Four and Germany.\footnote{For example, when we would analyze commodity prices and taking their volume as weights, the two figures of fixed (sum of volumes) and time-varying weights (daily volume) showed greater differences than in our case.} Comparing the two GDP weights, we conclude that proportions between countries when considering GDP at PPP per capita get closer to the situation when countries have equal position, thus equal effects on synchronization. It is shown in figures on the left of Fig.~\ref{fig:cohWeights} that the cohesion weighted by GDP per capita does not differ much in comparison to the one with equal weights.

We show that the time-varying weights, representing sizes of the economies, have substantial effect for business cycle synchronization measure. In our particular case, the absolute maximum in the difference between fixed and time-varying weights results is approximately 0.2, which comprises 10\% of the scale $[-1,1]$, see Fig.~\ref{fig:diffCohWeights} (right). The time varying weights correspond to precise moment of time, the information should be truly localized than that keeping weights constant. The results are shown in greater accordance with reality.

\subsubsection{Cohesion of the Visegrad Four and the EU}

We compute the multivariate relationship quantities for three groups of countries: the V4, the EU core, and the PIIGS countries. We analyze those groups individually as well as in combination.\footnote{By combination we mean the co-movement of all countries from both groups with respect to their weights.} %\footnote{The combination does not mean an analysis of co-movement of 2 groups but this provides the information about the co-movement of all countries of both groups with respect to their weights.}
We concentrate on the V4 itself and within the framework of the EU. Examining the synchronization of the PIIGS countries with the EU core, we directly compare already integrated countries with those in the process of integration.
\begin{figure}[ht]
  \begin{minipage}{0.47\textwidth}
    \includegraphics[width=\textwidth]{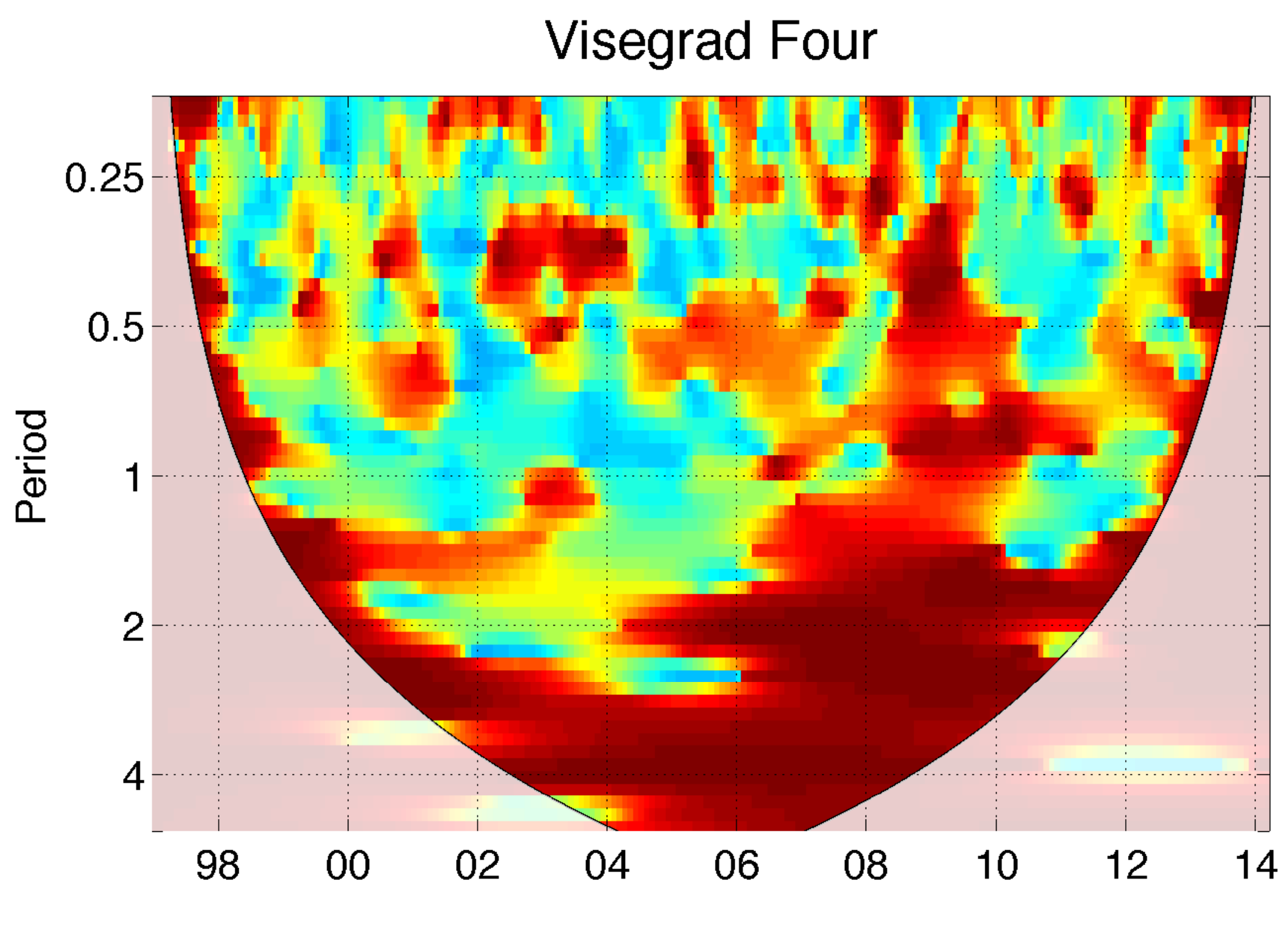}
  \end{minipage}
  \begin{minipage}{0.5\textwidth}
    \includegraphics[width=\textwidth]{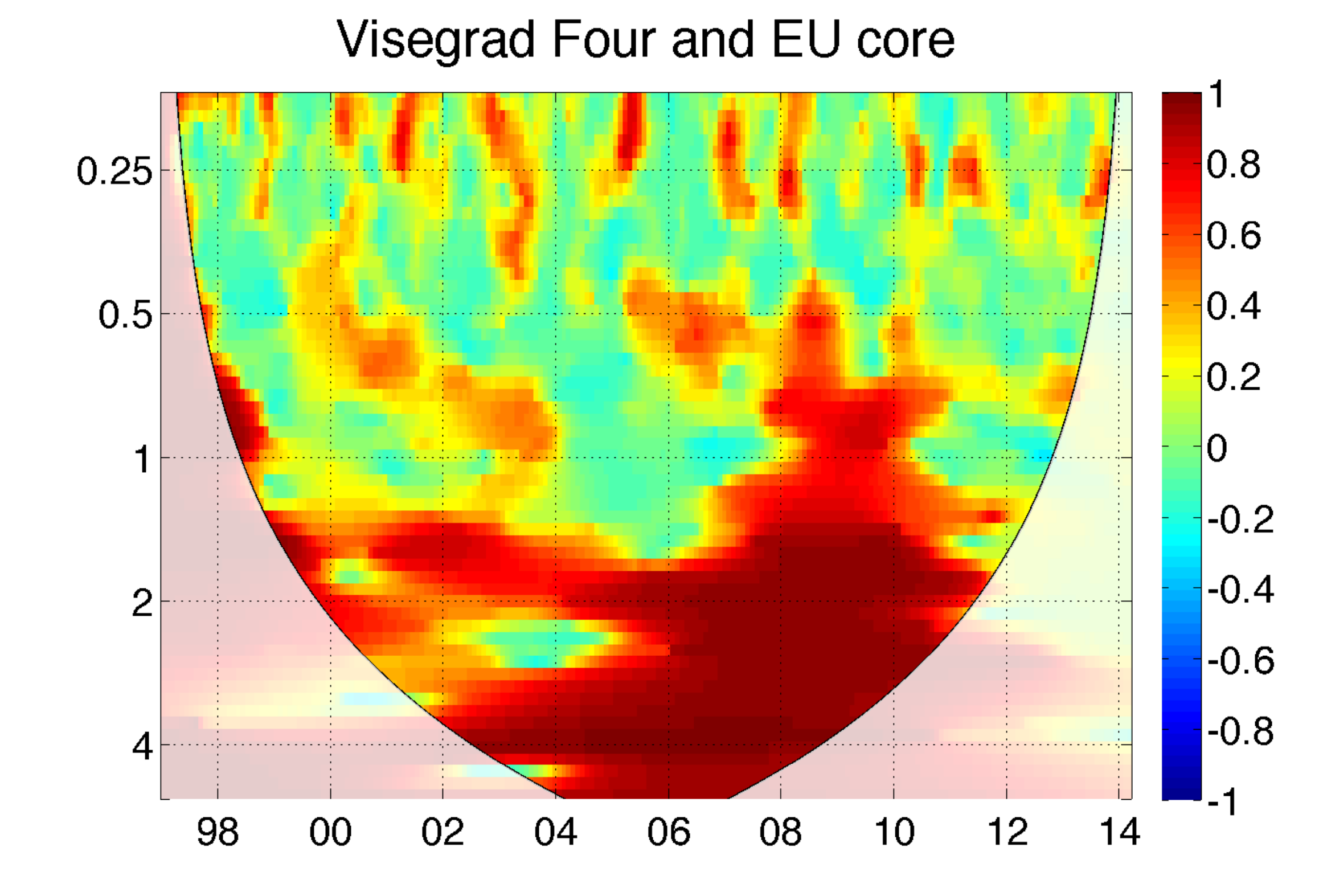}
  \end{minipage}\\
  \begin{minipage}{0.47\textwidth}
    \includegraphics[width=\textwidth]{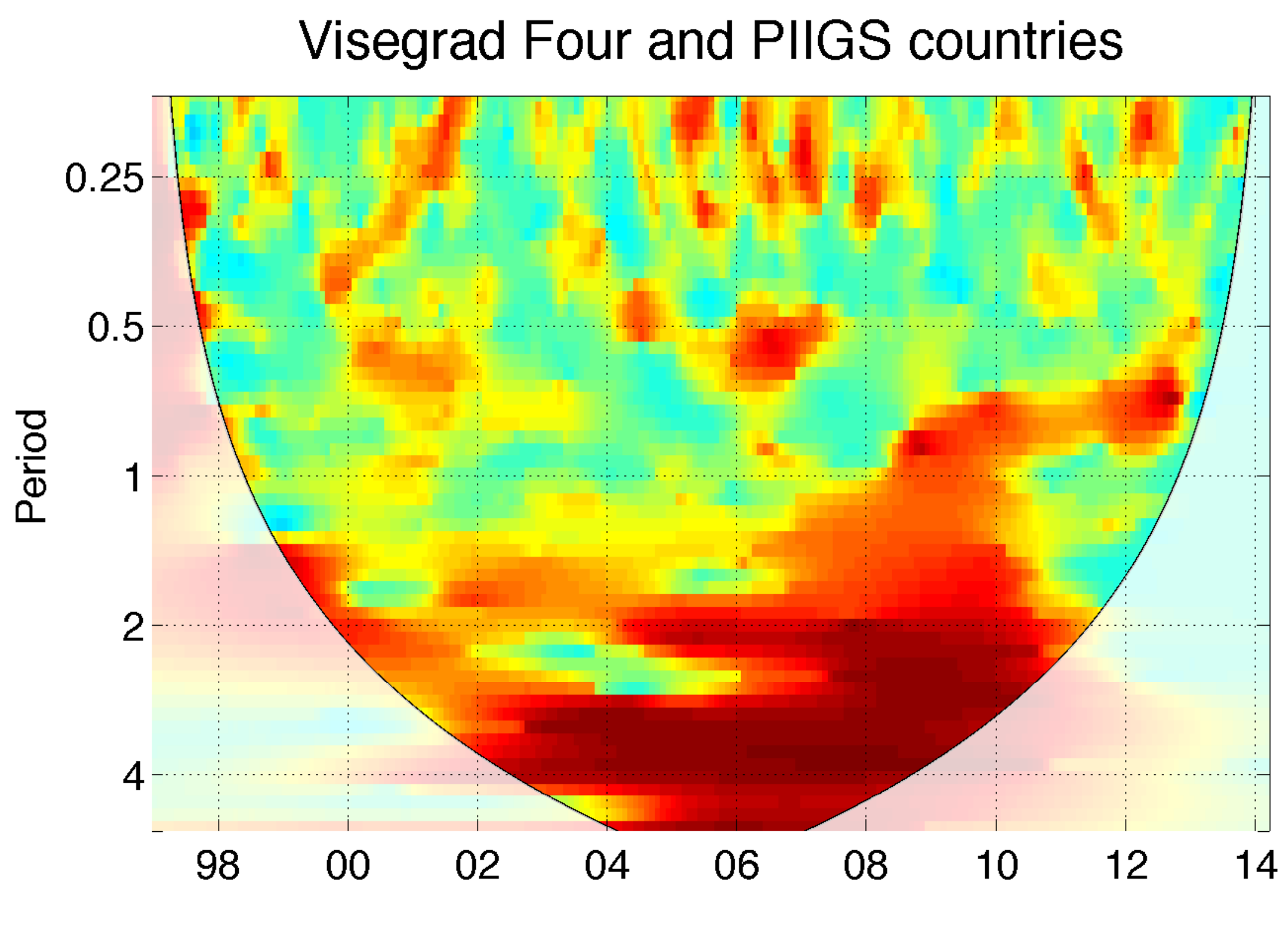}
  \end{minipage}
  \begin{minipage}{0.5\textwidth}
    \includegraphics[width=\textwidth]{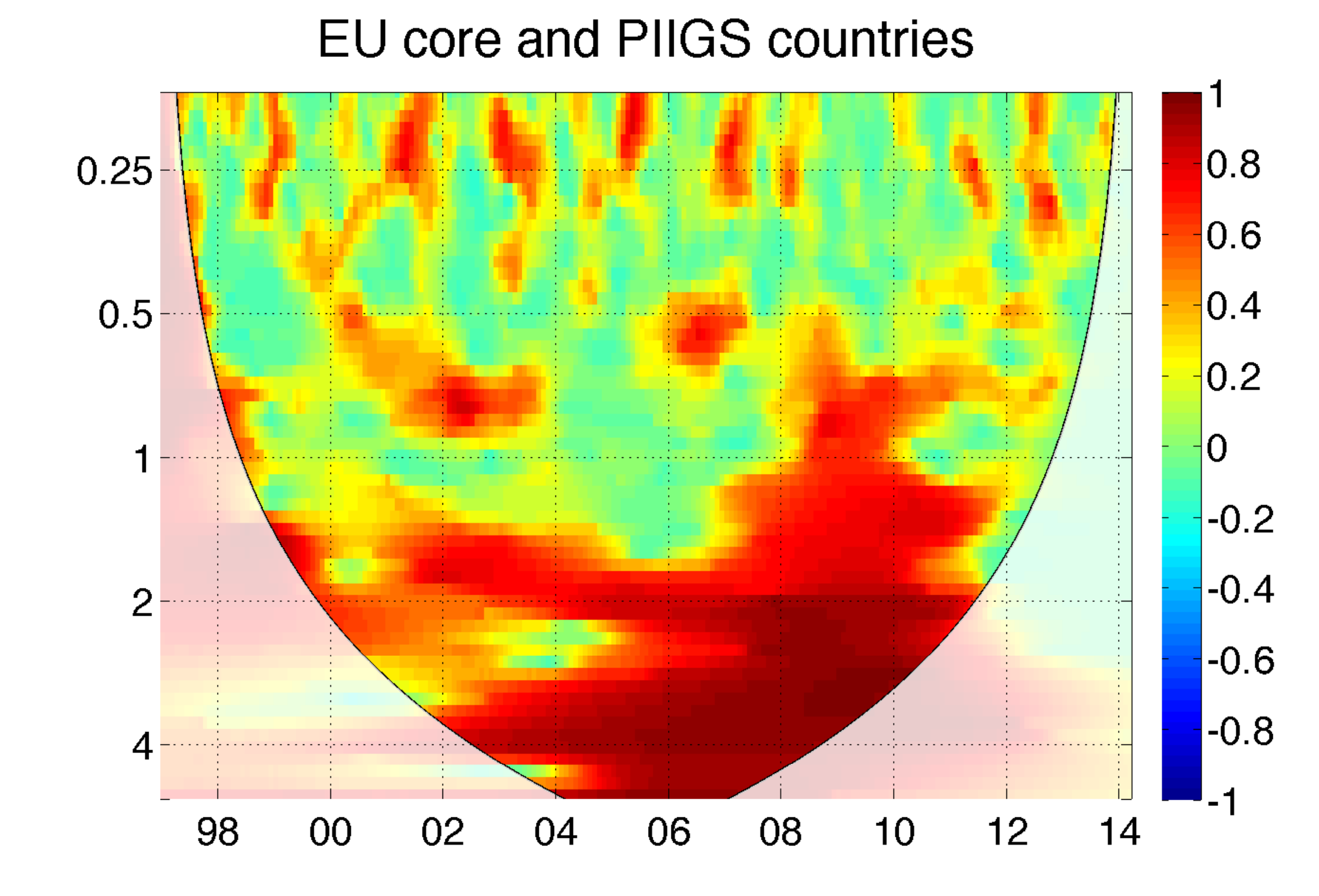}
  \end{minipage}
  \caption[Wavelet cohesion]{Wavelet cohesion of the Visegrad Four and parts of the EU. We use Gross Domestic Product as weights for this analysis. The shaded area is the cone of influence.}
  \label{fig:cohesionItself}
\end{figure}

\begin{figure}[ht]
  \begin{minipage}{0.31\textwidth}
    \includegraphics[width=\textwidth]{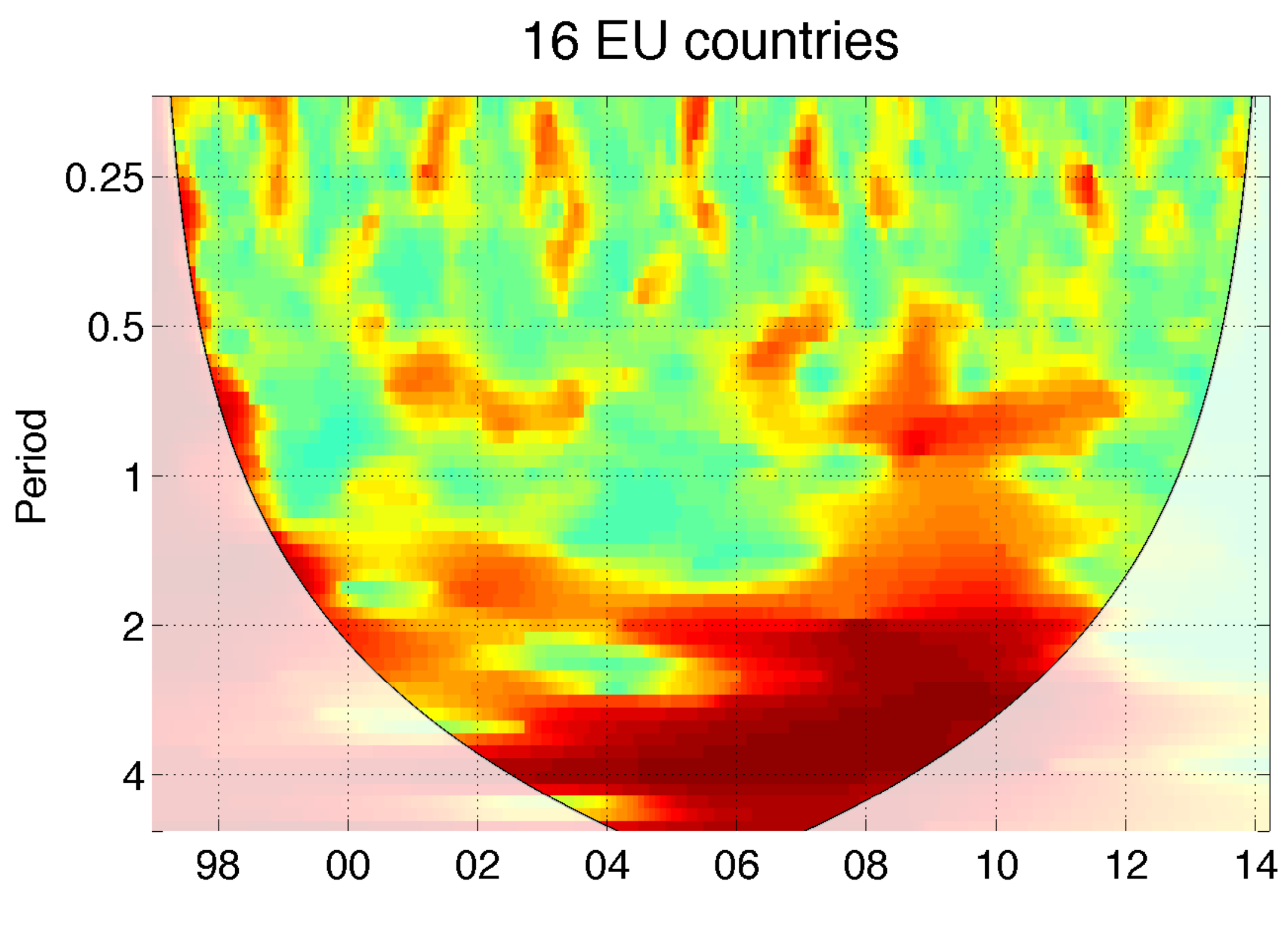}
  \end{minipage}
  \begin{minipage}{0.31\textwidth}
    \includegraphics[width=\textwidth]{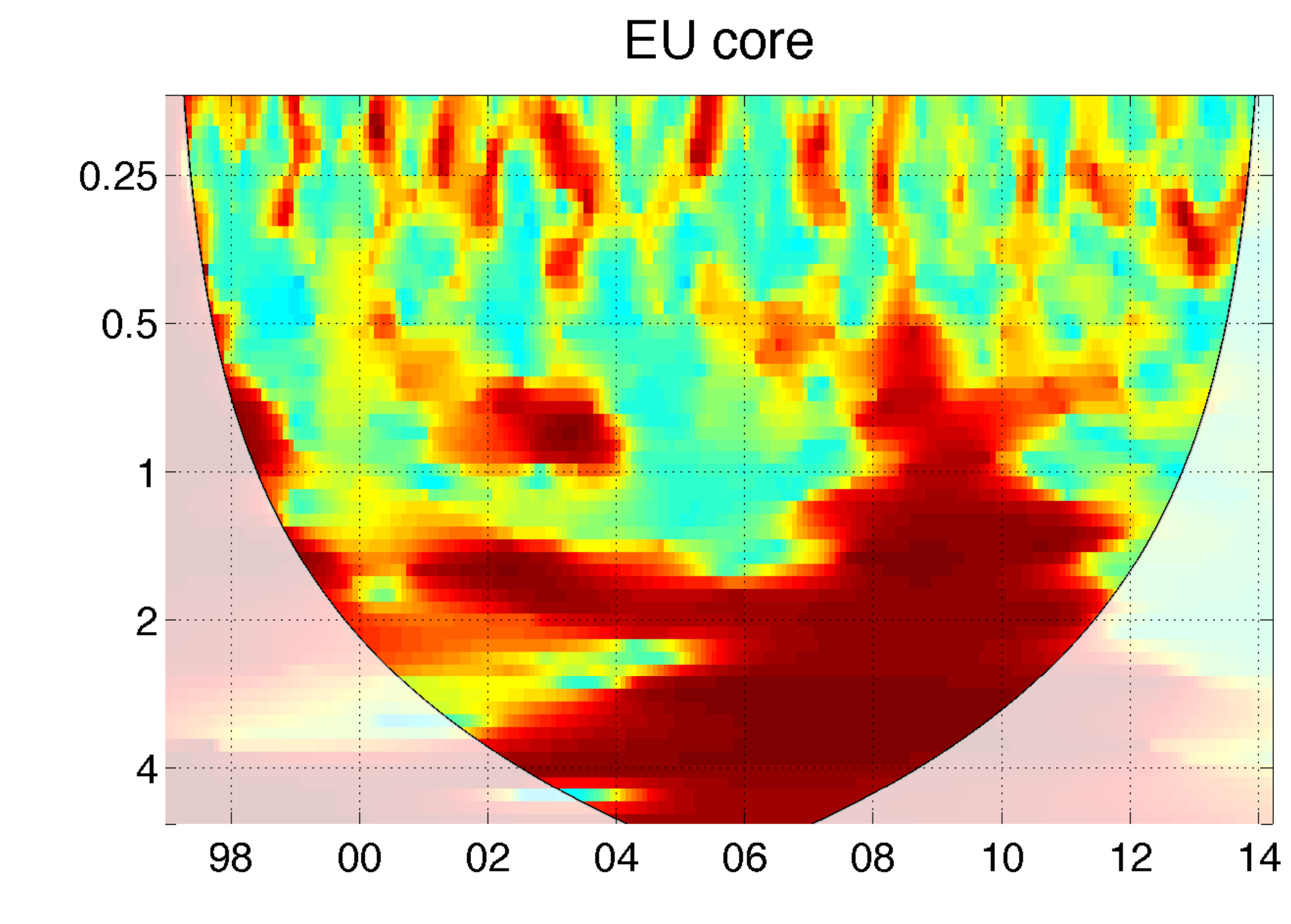}
  \end{minipage}
  \begin{minipage}{0.325\textwidth}
    \includegraphics[width=\textwidth]{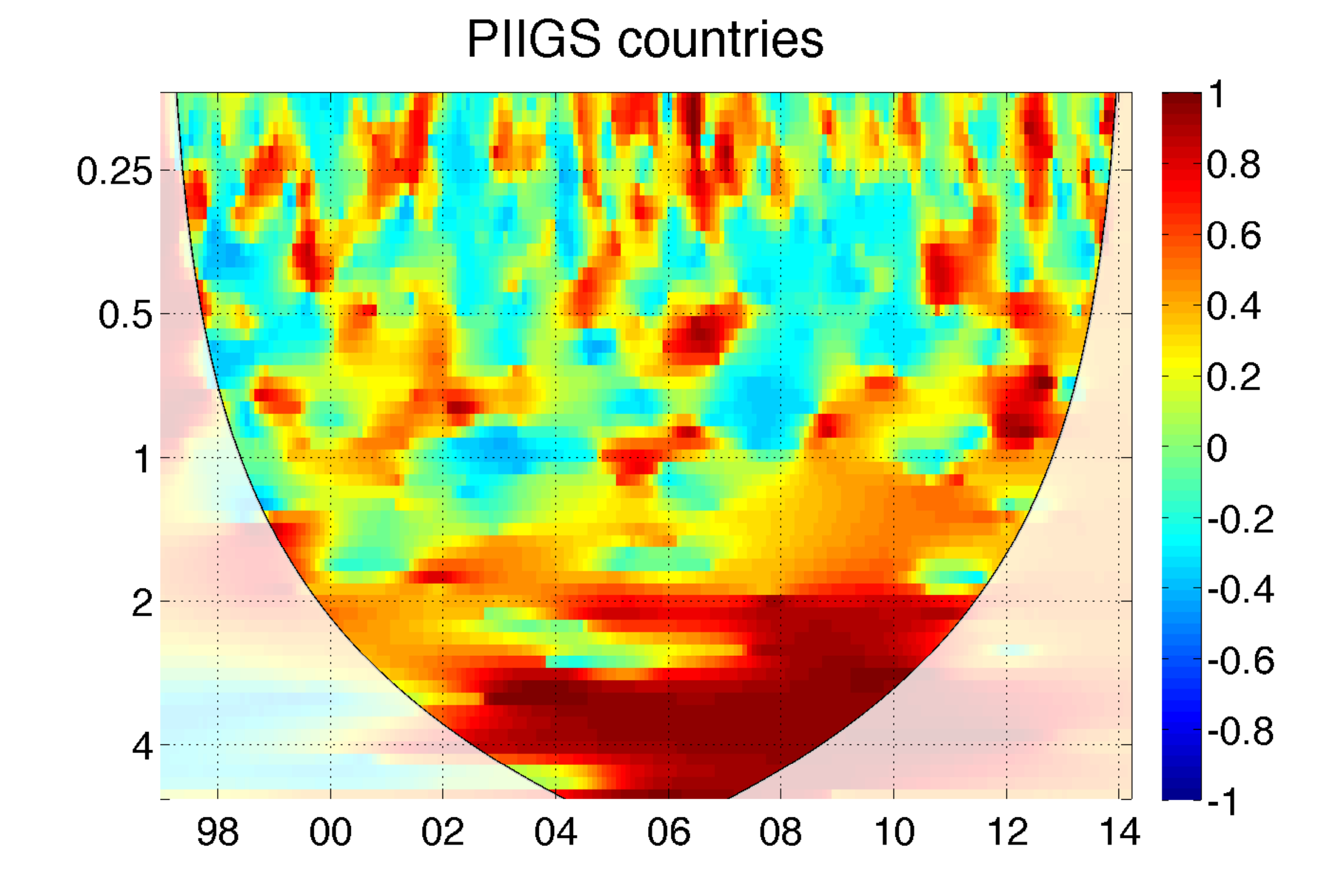}
  \end{minipage}
  \caption[Wavelet cohesion]{Wavelet cohesion of 16 European countries (left), EU core (center), and PIIGS countries (right). We use Gross Domestic Product as weights for this analysis. The shaded area is the cone of influence.}
  \label{fig:cohesionSolo}
\end{figure}

With regard to the V4, in Fig.~\ref{fig:cohesionItself}, we show that the degree of synchronization within this region is high in the long-term period. The strong co-movement of 2-4 year periods lasts from 1997 to 2014; beginning in 2008, it spreads over a 0.5-5 year period. The short-term synchronization, up to 1 year, provides some insights that countries often co-move at that frequency but only for short periods of time. This short-term fluctuation co-movement appears as well as for groups of EU core and PIIGS countries in Fig.~\ref{fig:cohesionSolo}, which may reflect common reactions to events in the markets.

Comparing the multiple relationships of V4 and EU core, EU core and PIIGS, and V4 and PIIGS countries in Fig.~\ref{fig:cohesionItself}, we observe similar patterns of co-movement over the long-term. This result of is in line with \citet{Rua:2012aa}, who find a large cohesion of the long-term dynamics. Regarding the Visegrad counties within the EU, in the period after 2000, the strongest relationship is among the V4 countries and the EU core compared to other groups.

Further, we investigate the relationship of all 16 selected countries and the sub-samples of the EU core and the PIIGS countries. Surprisingly, the analysis reveals low cohesion in the short-term, 0.5-1 year, although it is quite high in the long-term. Regarding the EU core, there is more commonality in short-term dynamics. Further for all groups, we note that the cohesion is high for long-term business cycles. In the second half of the sample, the relationship has increased for the shorter periods. With respect to the case of PIIGS counties, their cohesion appears to be along the same lines, but these countries do not exhibit any co-movement over the 1-2 year period. Notably, we have seen that the cohesion of the V4 with the EU core is stronger than that of the 16 European countries. For example, in Slovakia, there was a gradual increase of co-movement with Germany after accession and after the Euro adoption. In terms of synchronization, the V4 could benefit from joining the EMU, but we do not know the specific sources of the higher co-movement.

\section{Concluding remarks}
\label{sec:conclusion}

Business cycle synchronization is a central question of economic integration and thus it needs a rigorous examination. We have overcome the problems of traditional measures, such as operation in time or frequency domain only and of the necessity of stationary time series, by using wavelet methodology. In this paper, we have proposed the multivariate measure of co-movement with time-varying weights called wavelet cohesion. This wavelet-based measure allows for precise localization of information in time and frequency.

We have investigated the impact of V4 cooperation, which has one of its aims to converge faster towards the EU. We have found short but high co-movement for the first years of their cooperation until the economic turbulences of the late 90s. During the 1995-1999 period, their business cycles for 2-5 year periods show very low levels of synchronization.

Further, we have studied the business cycle synchronization of the V4 with Germany. The results confirmed some already known but interesting patterns. Slovakia's synchronization with the EU was poor before its accession to the EU but gets stronger after 2005, which supports the theory of the endogeneity of the OCA and the adoption of Euro. We have revealed that the highest coherence is between Germany and both the Czech Republic and Hungary beginning in 2000. By contrast, the degree of synchronization of the business cycles of Poland and Germany is the lowest among V4.

Employing a multivariate measure, we have uncovered relationships in both time and frequency domains for multiple time series. Regarding the V4, the EU core countries, and/or the PIIGS countries, we show that there is a very weak synchronization of short-term dynamics,less than 1 year, among those countries. Conversely, we have found that cohesion within the EU is high after 2000 in 2-5 year periods. Finally, we have found high co-movement of the long-term business cycles of the V4 and of the EU core countries over the whole sample. This supports that countries tend to have similar approaches to their long-term dynamics or policies.

\section*{References}

% \bibliography{notesBiblio}
\bibliography{viseCoheBib}
\bibliographystyle{chicago}

\end{document}